\documentstyle[aps,eqsecnum,epsf,axodraw]{revtex}
\def\gsim{\mathrel{\rlap {\raise.5ex\hbox{$ > $}}
{\lower.5ex\hbox{$\sim$}}}}
\def\lsim{\mathrel{\rlap {\raise.5ex\hbox{$ < $}}
{\lower.5ex\hbox{$\sim$}}}}

\newcommand{\be}{\begin{equation}}
\newcommand{\ee}{\end{equation}}
\newcommand{\bea}{\begin{eqnarray}}
\newcommand{\nn}{\nonumber}
\newcommand{\eea}{\end{eqnarray}}

\def\gappeq{\mathrel{\rlap {\raise.5ex\hbox{$>$}}
{\lower.5ex\hbox{$\sim$}}}}
 
\def\lappeq{\mathrel{\rlap{\raise.5ex\hbox{$<$}}
{\lower.5ex\hbox{$\sim$}}}}
 
\begin{document}
 
\begin{titlepage}
\begin{flushright}
ACT--1/98 \\
CTP-TAMU--7/98 \\
OUTP--98-15P \\
quant-ph/9802063 \\
\end{flushright}

\begin{centering}
\vspace{.1in}
{\large {\bf QUANTUM MECHANICS IN CELL MICROTUBULES: WILD IMAGINATION OR REALISTIC POSSIBILITY ?}} \\

\vspace{.2in}
{\bf N.E. Mavromatos$^{a}$} 
and 
{\bf D.V. Nanopoulos$^{b,c,d}$}

\vspace{2in} 

Based on a talk given by N.E.M. at the Workshop 
{\it Biophysics of the Cytoskeleton}, Banff Conference Center, Canada, August 18-22 1997 \\
To appear in {\it Advances in Structural Biology}

\end{centering}

\vspace{1.5in}
\begin{flushleft}
$^{a}$ P.P.A.R.C. Advanced Fellow, Department of Physics
(Theoretical Physics), University of Oxford, 1 Keble Road,
Oxford OX1 3NP, U.K.  \\
$^{b}$ Department of Physics, 
Texas A \& M University, College Station, TX 77843-4242, USA, \\
$^{c}$ Astroparticle Physics Group, Houston
Advanced Research Center (HARC), The Mitchell Campus,
Woodlands, TX 77381, USA, \\
$^{d}$ Academy of Athens, Chair of Theoretical Physics, 
Division of Natural Sciences, 28 Panepistimiou Avenue, 
Athens 10679, Greece. \\

\end{flushleft}

\end{titlepage} 

\newpage

\begin{centering}

\vspace{1cm}

{\bf Table of Contents} 

\end{centering} 

\noindent Abstract \dots \dots 1 

\noindent I. Introduction \dots \dots 1 - 6 

\noindent II. On the Rabi splitting in Atomic Physics and the quantum nature of electromagnetic Radiation \dots \dots 7 - 11 

\noindent IIA. Description of the Rabi splitting phenomenon \dots \dots 7 - 9 

\noindent IIB Rabi splitting and decoherence: experimental verification 
\dots \dots 10 - 11 

\noindent III. Microtubules as dielectric cavities and dissipationless energy transfer in the cell \dots \dots 11 - 22 

\noindent IIIA. Microscopic Mechanisms for the formation of coherent states in MT \dots \dots 11 - 13 

\noindent IIIB. Decoherence and dissipationless energy transfer \dots \dots 
13 - 22 

\noindent IV. Microtubules as Quantum Holograms \dots \dots 22 - 27

\noindent V. Outlook \dots \dots 27 - 29

\noindent Acknowledgements, References \dots \dots 30 - 32

\begin{centering}

{\bf ABSTRACT} \\
\vspace{.1in}
\end{centering}
{\small 
We focus  
on potential mechanisms for `energy-loss-free' transport 
along the cell microtubules, which could be considered
as realizations of Fr\"ohlich's ideas
on the r\^ole of solitons for superconductivity 
and/or biological matter. In particular, 
by representing the MT arrangements as {\it cavities},
we present a novel 
scenario on  
the formation of
macroscopic (or mesoscopic) 
quantum-coherent states, 
as a result of the (quantum-electromagnetic) 
interactions 
of the MT dimers with 
the surrounding 
molecules of the ordered water in the interior of the MT cylinders. 
We present 
some generic order of magnitude 
estimates of the decoherence time in a typical model for MT
dynamics. 
The r\^ole of (conjectured) ferroelectric properties of 
MT arrangements 
on the above quantum phenomena is emphasized. 
Based on these considerations, 
we also present a conjecture on the r\^ole of the 
MT in {\it holographic} information processing,
which resembles the situation encountered 
in internal source X-ray holography in atomic physics.}

\section{Introduction} 

It is known that for most aspects of biological 
systems the classical world proves sufficient in 
providing an accurate description. 
The macroscopic character of these systems, 
as well as their {\it open} nature, in the sense of thermal losses
due to interaction with the environment, usually at room temperatures, 
implies that, 
at the time scales relevant to most biological 
applications, the quantum nature of the system 
has disappeared in favour of the `collapse' to a classical ground state. 

However, there might still be some biological structures, 
somehow thermally isolated from the environment, which could 
sustain macroscopic quantum coherent states for some time, 
long enough so that certain functions are achieved
based on the properties of quantum mechanics. 
It is the purpose of this article to point out such a possibility,
in the context of energy transfer by microtubular structures inside the cell.

As is well known, {\it MicroTubules} (MT) appear to be 
one of the most fundamental structures of the 
interior of living cells (Dustin 1984, Engleborghs 1992).
These are paracrystalline cytoskeletal
structures which seem to play a fundamental r\^ole 
for the cell {\it mitosis}.
It is also believed
that they play an important r\^ole 
for the transfer of electric signals
and, more general, of energy 
in the cell. 

In this latter respect it should be mentioned 
that energy transfer across the cells,  
without dissipation,  
had been conjectured to occur in 
biological matter 
by 
Fr\"ohlich (Frohlich 1986), already some time ago. 
The phenomenon conjectured by Fr\"ohlich was based 
on his one-dimensional superconductivity model:
in one dimensional electron systems with holes, 
the formation 
of solitonic structures due to electron-hole pairing
results 
in the transfer of electric 
current without dissipation.
In a similar manner, Fr\"ohlich conjectured 
that energy in biological matter could be transfered 
without dissipation, if appropriate solitonic 
structures are formed inside the cells.
This idea has lead theorists to construct 
various models for the energy transfer 
across the cell, based on the formation of kink 
classical solutions (Lal 1985). 

In the early works no specific microscopic models had been 
considered (Lal 1985).
Recently, however, after the identification of 
the MT as one of the most important structures of the cell,
both functionally and structurally, 
a model for their dynamics 
has been presented (Satari\'c {\it et al.} 1993), 
in which the formation of solitonic structures,  
and their r\^ole in energy transfer across the MT, is discussed
in terms of classical physics.   
Mavromatos and Nanopoulos (Mavromatos and Nanopoulos 1997) 
have considered the {\it quantum aspects}
of this one-dimensional model, and argued on the consistent 
quantization of the soliton solutions, as well as the fact that 
such semiclassical solutions may emerge as a result of `decoherence'
due to environmental entanglement, according to recent 
ideas (Zurek 1991, 1981, Caldeira and Leggett 1983).  

The basic assumption of the model of [Mavromatos and Nanopoulos]
was that the fundamental structures
in the MT (more specifically of the brain MT) 
are Ising spin chains (one-space-dimensional structures).
The interaction of each chain with the neighboring
chains and the surrounding water environment 
had been mimicked by suitable potential terms in the 
one-dimensional Hamiltonian.
The semi-microscopic lattice model 
used to describe the dynamics of such one-dimensional 
sub-structures 
was the ferroelectric distortive spin chain
model of [Satari\'c {\it et al.}]. The basic field-theoretic degree
of freedom is the displacement field, defined 
by the projection of the 
difference of the electric-dipole vectors on the axis of the MT
cylinders. The basic assumption of the model 
is that the electric dipole vectors 
of the dimers
in the two conformations 
$\alpha, \beta$ differ by a constant angle $\theta$ in their orientation.

The one-dimensional nature of these fundamental
building blocks, then, opens up the way for a mathematical
formulation of the chain as a {\it completely intergrable} 
field theory model, characterized by an infinite-number
of conservation laws. The latter are associated with 
global excitation modes, 
completely {\it delocalised} in the dimer chain space-time.
This integrability structure 
proves sufficient
in providing a satisfactory solution
to memory coding and capacity
(Mavromatos and Nanopoulos 1997). Such features
might turn out to be important
for a model of the brain as a {\it quantum computer}.

The coupling of the MT chain to its `environment'
may lead, according 
to the analysis of [Mavromatos and Nanopoulos 1997],  
to the formation of {\it macroscopic quantum coherent 
solitonic} states. 
The latter arise from  {\it canonical quantization} of 
classical soliton (kink) solutions, derived from a 
certain {\it Hamiltonian}. It should be noticed that 
in the original model of [Sataric {\it et al.}],
the friction term,  describing the coupling 
of the dimers with the surrounding water molecules,
is put in phenomenologically as a (non-lagrangian) frctional force term.
However, in the context of the quantum integrable 
model of [Mavromatos and Nanopoulos 1997], the interaction 
of the dimers with the environment of water molecules
is given a `more microscopic' description, 
in terms of an induced `metric' in the one-dimensional 
space of the dimer chains. The 
result of such a representation, apart from the complete intergrability
of the model, is the fact that  
kink soliton solutions for the electric-dipole displacement field 
are obtained as solutions
of a Hamiltonian problem. 
This implies that they can 
be quantized {\it canonically} (Tsue and Fujiwara 1991), 
leading to macroscopic quantum coherent states.  
It can be shown that the kink solutions of [Sataric {\it et al.}]
are connected to the (classical part) 
of the above quantized solitons by appropriate field redefinitions. 
In this context, the details of the environment
would play a crucial r\^ole in determining the 
time scales over which such coherent states could be 
sustained {\it before collapsing to classical ground states}. 

Such analyses have yield
important results in the past
(Ellis {\it et al.} 1984, Ellis, Mohanty, Nanopoulos 1989, Zurek 1991, 
Gisin and Percival 1993), 
concerning the passage from the quantum to classical 
world. In particular, in [Zurek 1991] it was suggested
that quantum coherent states may appear as a result of
decoherence, depending on the nature of the environment (Albrecht 1992). 
Such states are minimum entropy/uncertainty states, which propagate 
`almost classically' in time, in the sense that their shape 
is retained during evolution. 
Such `almost classical' states 
are termed 
`pointer states' by Zurek.  
In view of the results of 
[ Gisin and Percival], concerning
the localization of the wave function
in open (stochastic) quantum-mechanical systems, the emergence of 
pointer states, as a result of decoherence, 
may be interpreted as implying 
that the localization process of the state vector had stopped
at a stage where 
it is not complete, but is such that the resulting (minimum-entropy) 
state
is least susceptible to the effects of the environment. 

It will be instructive to expose the reader to some formalism,
which eventually can be used to obtain some numbers,
in connection with the application of the above ideas
to the physics of MT. 
Consider an open system, described by a Hamiltonian $H$,
in interaction with a quantum-mechanical environment, described 
generically by  annihilation and creation operators $B_m$, $B^\dagger_m$.
For simplicity we assume a specific form of environmental entanglement,
which is such that the total probability and energy 
are conserved on the average.
This sort of condition are assumed to be met in the biological 
systems of interest [Mavromatos and Nanopoulos]. Such 
environmental entanglement is described by the so-called 
Lindblad formalism (Lindblad 1976) for the evolution equation 
of the density matrix of the system:
\be
{\dot \rho } \equiv \partial _t \rho
= i [\rho, H] - \sum _m
\{ B^\dagger_m B_m, \rho  \}_{+}
+ 2 \sum _m  B_m\rho  B^\dagger _m
\label{bloch}
\ee
which is a typical linear evolution equation for $\rho$, used in most 
Markov processes in open system quantum mechanics (Gorini {\it et al.} 1976).

We now remark that the density matrix is expressed in terms 
of the state verctor $|\Psi >$ via 
\be
  \rho = {\rm Tr}_{\cal M} |\Psi > < \Psi |
\label{neumanform}
\ee
with the trace being taken over en ensemble of theories ${\cal M}$.

The time evolution equation for the state vector, corresponding to 
(\ref{bloch}) acquires the {\it stochastic} Ito form,
which in an appropriate differential form reads (Gisin and Percival ):
\bea
&~&|d\Psi > =-\frac{i}{\hbar} H |\Psi >
+ \sum _m (<B^\dagger _m >_{\Psi} B_m - \frac{1}{2}
B^\dagger _m B_m - \nn \\
-&~&\frac{1}{2} <B^\dagger _m>_{\Psi}
<B_m>_{\Psi })~|\Psi > dt +
\sum _m
(B_m - <B_m>_{\Psi})~|\Psi > d\xi _m
\label{ito}
\eea
where 
$<\dots>_{\Psi}$ denote averages with respect to the
state vector $|\Psi >$, and $d\xi _m$ are complex differential
random matrices, associated with white noise Wiener or
Brownian processes. 
As regards eq. (\ref{ito}) attention is drawn
to a recent formal derivation (Saito and Aremitsu 1993) of this equation
from Umezawa's thermo-field dynamics model~ (Umezawa 1993) ,
which is a quantum field theory model for open systems
interacting with the environment.

Within the stochastic framework implied by (\ref{ito}) it
can, then, be shown
that under some not too resrtictive
assumptions for the Hamiltonian operator
of the system, localization of $|\Psi >$ within
a state-space channel, $k$, will always occur, as a result of
environmental entanglement.
To prove this
formally, one constructs a quantity that serves as a
`measure' for the delocalization
of the state vector, and examines its temporal
evolution. This is given by the so-called
`quantum dispersion entropy'~(Gisin and Percival)
defined as
\be
  {\cal K} \equiv -\sum _k <P_k>_{\Psi}ln~<P_k>_{\Psi}
\label{dispentr}
\ee
where $P_k$ is a projection operator in the `channel' $k$
of the state space of the system. 
This entropy is shown to decrease in situations where
(\ref{ito}) applies, under some
assumptions about the commutativity of the
Hamiltonian of the system with $P_k$, which implies
that $H$ can always be written in a block diagonal form.
The result for the rate of change of the dispersion
entropy is then
\be
  \frac{d}{d t} ({\cal M}{\cal K}) = -\sum _k \frac{1 - <P_k>_{\Psi}}
{<P_k>_{\Psi}} R_k \le 0
\label{rate}
\ee
where $R_k$ are the (positive semi-definite)
{\it effective interaction rates} in channel $k$,
defined as (Gisin and Percival) 
\be
    R_k \equiv \sum _j |<L_{kj}>_{\Psi}|^2
\label{eir}
\ee
with $L_{kj} \equiv P_k L_j P_k $ denoting the projection
of the environment operators in a given channel.

The presence of `pointer states' 
amounts essentially in the fact that in
such a case the state-vector localization procedure 
stops before it is complete. The resulting ground state is a 
coherent quantum state, with a finite life time. 
Eventually of course, especially in biological systems, temperature 
effects, due to the lack of complete thermal isolation, induce a
collapse to a completely classical state. Our point in this article 
is to present a scenario for the emergence of such coherent 
states in MT chains, and estimate the life time and decoherence
time scales of the resulting quantum coherent `pointer ' state. 

Notice that 
not all environments admit such pointer states (Albrecht).
Here lies one of the advantages of 
viewing the MT system of chains
as a completely integrable Liouville-string theory.
The model possesses a pointer basis (Ellis, Mavromatos, Nanopoulos 1994),
which in [Mavromatos, Nanopoulos 1997] has been identified with the 
quantized (via squeezed-coherent states) 
soliton solutions of [Satari\'c {\it et al.}].
Another advantage lies on the 
fact that energy is {\it conserved} on the average
in the model, despite the 
environmental entanglement. This is for purely 
stringy reasons pertaining to the 
Liouville approach. The reader may find details
in [Ellis, Mavromatos, Nanopoulos 1992, 1993].
Physically this means that such a formalism
is capable of describing {\it energy-loss-free} transport 
in biological cells, thereby providing a realization
of Fr\"ohlich's ideas. 

In this article we focus our attention 
on the {\it microscopic nature} 
of such coherent states, and try to understand their emergence 
by as much `conventional' physics arguments as possible. By 
`conventional' we mean scenaria based on 
{\it quantum electrodynamics},
which are the major interactions expected to dominate 
at the energy scales of the 
MT. We make an attempt to pick up possible sources of 
decoherence, consistent with the above-mentioned
energy-loss-free transfer scenario in biological cells, and 
estimate the associated time scales.

The {\it quantum nature } of the MT network could result
from the {\it assumption} that each dimer finds itself in a
superposition of $\alpha$ and $\beta$ conformations 
(Hameroff and Penrose 1996)~\footnote{The 
terminology `conformations' is used here strictly in the 
sense of representing the `localisation' of the unpaired 
charges in one of the tubulin monomers (`hydrophobic pockets').
It should not be confused with the terminology 
used in biology, which implies also additional electrochemical processes
in the interior of the protein monomers. For our purposes here, 
the terminology `$\alpha$ and $\beta$ conformations' 
refers to two eigenstates of the quantum 
mechanical system of the unpaired charges.} 

Viewed as a {\it two-state} quantum mechanical 
system, the MT tubulin dimers are {\it assumed} (Jibu {\it et al.},
Hameroff and Penrose )
to couple to such conformational changes with
$10^{-9}-10^{-11} {\rm sec}$ transitions, corresponding to an
angular frequency $     \omega \sim{\cal O}( 10^{10})
-{\cal O}(10^{12})~{\rm Hz}$. In the present work we 
assume the upper bound of this frequency range 
to represent (in order of magnitude) the characteristic frequency 
of the dimers, viewed as a two-state quantum-mechanical system: 
\be
     \omega _0 \sim {\cal O}(10^{12})~{\rm Hz} 
\label{frequency2}
\ee

We emphasize that in our scenario, 
an important r\^ole for the existence of the 
solitonic structures in the dimer chains 
is played by the ordered
water in the interior of the MT. The importance of the water 
in the interior of the MT has been already pointed out by S. Hameroff,
in connection with {\it classical } 
information processing (Hameroff 1974, 1987; Hameroff {\it et al.} 1986).
In our article we present a scenario for the formation 
of {\it quantum coherent modes} in the water 
of the MT, inspired by earlier 
suggestions on `laser-like' behaviour of the water (E. Del Giudice,
Preparata, Vitiello 1988, E. Del Giudice {\it et al.} 1985, 1986;
Jibu {\it et al.} 1994) arising from the interaction 
of the electric dipole moments of the water molecules .
with {\it selected} modes 
of the {\it quantized} elelctromagnetic radiation. 

In [Mavromatos and Nanopoulos 1997b] such coherent modes 
have also been argued to 
couple with the unpaired 
electrons of the MT dimers, resulting 
in {\it electric dipole wave quantum coherent modes}
in the dimer chains, which from a formal point of view have been argued 
to 
represent canonically quantized soliton waves in the model 
of [Satari\'c {\it et al.}]. 
Such a quantum coherent ordering 
may extend over the mesoscopic scales of the MT cylindrical arrangeents,
involving all the 13 protofilaments. 
Such modes are characterized 
by infinite dimensional symmetries (quantum integrability), 
corresponding to global excitations modes.
The excitation of the respective {\it quantum numbers}, thus, 
corresponds to a sort of {\it infinite-dimensional} coding, which 
might be able of offering a solution to the 
problem of memory capacity, should such considerations
apply to the brain cell MT (Mavromatos and Nanopoulos 1997a). 

From the physical point of view, it was argued in 
[Mavromatos and Nanopoulos 1997b]
that the 
coupling of the dimer chains with the coherent modes 
in the water interior of the MT
would result in the 
so-called 
Vacuum Field Rabi 
Splitting (VFRS) (Sanchez-Mondragon {\it at al.} 1983), 
a sort of dynamical Stark effect occuring 
in the emission or absorption spectra of the dimers 
as a result of the vaccum quantum fluctuations. 
This phenomenon is characteristic of the behaviour of 
atoms inside quantum electrodynamical 
cavities, the r\^ole of which is played in our work 
by the MT cylindrical structures
themselves. 

The atomic Rabi Splitting  is a consequence  
of the coupling 
of the atoms with 
the coherent modes of electromagnetic radiation inside the cavity. 
It is  
considered by many as a `proof' of 
the quantum nature of the electromagnetic radiation
(Sanchez-Mondragon {\it at al.}).
In our work, we consider this phenomenon, if true, 
as a manifestation of the quantum mechanical nature of the 
MT arrangement inside the cell. 

In our scenario, 
the MT arrangements inside the cell act as fairly 
isolated electromagnetic {\it cavities}. The ordered-water
molecules in their interior, then, provides an environment,  
necessary to form coherent quantum modes -
dipole quanta~(E. Del Giudice {\it et al.} 1988,
E. Del Giudice {\it et al.} 1985, 1986;
Jibu {\it et al.} 1994),
whose coupling 
with the dimers results in VFRS in the respective absorption 
spectra. Due to dissipation from the cavity walls, there is 
decoherence of the combined dimer-cavity system coherent state. 
The resulting decoherence time scales are estimated,
for various sources of envirnoments.

In order for the quantum coherent states to play a r\^ole 
in dissipationless energy transfer, 
the decoherence time must be larger than the 
the scale 
required for energy transfer in MT.  
In the model of Satari\'c {\it et al.},
for a moderately long MT,
of length $L \sim 10^{-6}$ m, 
the kink solitons 
transfer energy along the MT chains 
in a time scale of order: 
\be
t_{kink} = {\cal O}(5 \times 10^{-7}~sec)
\label{FS}
\ee
The main conclusion of our analysis 
in this work is that 
the decoherence time scale 
can indeed be larger than (\ref{FS}), provided that 
the damping time scales of the MT cavities are
$T_r \ge 10^{-4}-10^{-5} {\rm sec}$. As we shall argue, such 
numbers are not impossible to be met in Nature. 

If this picture turns out to be true, 
one can then use MT as biological 
cavities, and perform experiments 
analogous to those used in Quantum Optics, involving 
Rydberg atoms in electromagnetic cavities~(Bernardot {\it et al.} 1992,
Brune {\it et al.} 1990).
The hope is that such quantum mechanical aspects, if true, might 
be relevant for a better understanding of the physics of 
assemblies of 
cell microtubules, and in particular brain-cell MT, 
and their possible r\^ole in {\it quantum computation}
(Penrose 1994, Nanopoulos 1995, 
Mavromatos and Nanopoulos 1997a,b, Hameroff and Penrose 1996).

\section{On the Rabi Splitting in Atomic Physics and the Quantum Nature of 
Electromagnetic Radiation} 

\subsection{Description of the Rabi splitting phenomenon}

For instructive purposes, 
in this section we shall recapitulate briefly
the VFRS in atomic physics, which will be the basis 
for the subsequent discussion involving MT networks. 
This phenomenon has been 
predicted for the emission spectra of 
atoms inside electromagnetic cavities (Sanchez-Mondragon {\it et al.}), 
in an attempt
to understand the {\it quantized nature} of the electromagnetic
radiation.   
The basic principle underlying the phenomenon is 
that, in the presence of an interaction among two oscillators
in resonance, 
the frequency degeneracy is {\it removed} by an amount proportional 
to the strength of the coupling. In the cavity $QED$ case, 
one oscillator consists of 
a small collection of $N$ atoms, whilst the other is 
a resonant mode of a high-$Q$(uality) cavity~\footnote{The quantity $Q$ is 
defined as the ratio of the 
stored-energy to the energy-loss per period (Haroche and Raimond 1994),
by making the analogy with a damped harmonic oscillator.}.  
Immediately after the suggestion of Sanchez-Mondragon {\it et al.}
a similar phenomenon has been predicted for absorption spectra
of atoms in cavities~(Agarwal 1984). 

By now, the situation has been verified experimentally
on a number of occasions (Bernardot {\it et al.} 1992). In such experiments
one excites the coupled atom-cavity 
system by a tuneable field probe. The excitation is then found resonant 
{\it not} at  the `bare' atom or cavity frequencies but 
at the {\it split} frequencies of the `dressed' atom-field system. 
The spliting is enhanced for collections of atoms.
For instance, as we shall review below, 
for a system of $N$ atoms, the split is predicted to be~(Agarwal):
\be
     {\rm Rabi~splitting}=2\lambda \sqrt{N} 
\qquad 2\lambda={\rm Rabi~splitting~of~a~single~atom}
\label{rabienhanced}
\ee
Despite its theoretical prediction by means of 
quantum mechanical oscillator systems coupled with 
a {\it quantized} radiation field mode in a cavity, 
at present there seems
to be still a {\it debate} 
on the nature of the phenomenon: (i) the dominant opinion
is that the Rabi splitting is a manifestation of the {\it quantum
nature} of the electromagnetic radiation (cavity field), and is 
caused as a result of an {\it entanglement} between the atom and the cavity 
coherent modes of radiation. It is a sort of Stark effect, 
but here it occurs in the 
absence of an external field. This `dynamical Stark effect' 
is responsible for a splitting of the resonant lines 
of the atoms by an amount proportional to the collective atomic-dipole 
amplitude. 
(ii) there is however a dual 
interpretation~(Y. Zhu {\it et al.} 1990), which claims 
that the splitting can be observed in optical cavities as well,
and it is simply a result of {\it classical} wave mechanics inside the 
cavity, where the atomic sample behaves as a {\it refractive medium} 
with a {\it complex} 
index, which splits the cavity mode into two components.

Irrespectively of this second classical interpretation, one {\it cannot deny}
the presence of the phenomenon 
in entangled atom-quantum-coherent mode systems.
This is the point of view we shall be taking in this work, 
in connection with our picture of viewing MT filled with ordered water
as cavities. We shall try to make specific experimental
predictions that could shed light in the formation 
of quantum coherent states, and their eventual decoherence.
As mentioned above, the latter  
could be due to the interaction of the dimer unpaired spins
(playing the r\^ole of the atoms in the Rabi experiments) 
with the ordered-water coherent modes (playing the r\^ole 
of the cavity fields). Possible scenaria for the origin of such cavity
coherent modes will be described below. 

Let us first recapitulate briefly the theoretical 
basis of the Rabi-splitting phenomenon, which will allow the non-expert reader
to assess the situation better. We shall present 
the phenomenon from a point of view that will help us 
transcribe it directly to the MT case. 
Consider an atom of a frequency $\omega_0$ 
in interaction with a single coherent
mode of electromagnetic radiation field of frequency $\omega$.
The relevant 
Hamiltonian is:
\be
H=\hbar\omega_0\sum_{i} S_i^z + \hbar \omega a^\dagger a 
+ \sum_{i} (\hbar \lambda S_i^+ a + H.C.)
\label{hamrabi}
\ee
where $a^\dagger,a$ are the creation and annihilation 
cavity radiation field modes, 
$S_i^z$,$S_i^{\pm}$ are the usual spin-$\frac{1}{2}$ 
operators, and $\lambda$ is the atom-field coupling. 
The atom-field system is not an isolated system, since there is 
{\it dissipation} due to the interaction of the system with the 
surrounding world. An important source of dissipation is 
the leakage of photons from the cavity at some rate $\kappa$.
If the rate of dissipation is not too big, then 
a quantum coherent state can be formed, which would allow 
the observation of the vacuum-field Rabi oscillations. 
The density matrix $\rho$ of the atom-field system obeys a Markov-type 
master equation for the evolution 
in time $t$~(Agarwal):
\be
\partial_t \rho =-\frac{i}{\hbar} [H, \rho] - \kappa (a^\dagger
a \rho - 2 a \rho a^\dagger + \rho a^\dagger a )
\label{markovrabi}
\ee
The limit $\kappa << \lambda \sqrt{N}$ guarantees
the possibility of the formation of a quantum coherent 
state, i.e. this limit describes environments that 
are weakly coupled to the system, and therefore the 
decoherence times (see below) are very long. 
In this limit one can concentrate on the off-diagonal elements
of the density matrix, and make the following 
(`secular') approximation for their evolution~(Agarwal):
\be
\partial _t \rho _{ij} =-\frac{i}{\hbar} 
(E_{i} - E_{j} )\rho_{ij} - \Gamma _{ij} \rho_{ij} 
\label{secular}
\ee
where $\Gamma _{ij}$ denotes the damping factor, pertaining 
to the weak coupling of the atom-field system 
to the environment.  The analysis of [Agarwal]
pertained to the evaluation of the susceptibility tensor
of the system, $\chi_{\alpha\beta}$,
which can be calculated by considering 
its interaction of the system with 
an external field of frequency $\Omega$. 
The absorption spectrum is proportional to 
${\rm Im}\chi (\Omega) $. A standard quantum-mechanical 
computation 
yields the result:
\bea
 &~&{\rm Im}\chi(\Omega) ={\rm cos}^2\theta
\frac{\Gamma _-/\pi}{\Gamma _-^2 + 
\{ \Omega - \omega_0 + \Delta/2 -\frac{1}{2}
(\Delta ^2 + 4 N \lambda ^2 )^{1/2}\}^2}
+ \nonumber \\
&~&{\rm sin}^2\theta \frac{\Gamma _+/\pi}{\Gamma _+^2 + 
\{ \Omega - \omega_0 + \Delta/2 + \frac{1}{2}
(\Delta ^2 + 4 N \lambda ^2 )^{1/2}\}^2}
\label{suscept}
\eea
with $\Delta \equiv \omega_0 - \omega$.
In the above expression the damping factors $\Gamma _{\pm}$ 
represent the damping in the equation of motion for the 
off-diagonal element of the density matrix $<\Psi_0|\rho|\Psi_{\pm}^{S,C}>$
where $\Psi_{\pm}$ are eigenfucntions of $H$, classified by the eigenvalues 
of the operators $S^2$, and $S^z+a^\dagger a \equiv C$, in 
particular:
$S=N/2$, and $C=1 -N/2$.

The expression (\ref{suscept}) summarizes the 
effect of Rabi-vacuum splitting in absorption spectra of atoms: 
there is a doublet structure (splitting) of the absorption spectrum
with peaks at:
\be
\Omega = \omega _0 - \Delta/2 \pm \frac{1}{2}( \Delta ^2 + 
4 N \lambda ^2 )^{1/2}
\label{rabiabs}
\ee
For resonant cavities the splitting occurs with equal weights
\be
  \Omega = \omega_0 \pm \lambda \sqrt{N} 
\label{rabisplitting}
\ee
Notice here the {\it enhancement} in the effect 
for multi-atom systems $N >> 1$. 
This is the `Vacuum Field Rabi Splitting phenomenon', predicted
in emission spectra in [Sanchez-Mondragon {\it et al.} ]. As we have already
mentioned, the above derivation
pertains to absorption spectra, where 
the situation is formally much simpler (Agarwal).
It is also this latter case that is of interest to us for the purposes
of this work. 

The quantity  $2\lambda \sqrt{N}$ is called the `Rabi frequency'. 
{}From the emission-spectrum theoretical analysis 
an estimate of $\lambda$ may 
be inferrred which involves 
the matrix element, ${\underline d}$, of atomic electric dipole 
between the energy state
of the two-level atom~(Sanchez-Mondragon {\it et al.}): 
\be 
   \lambda = \frac{E_{vac}{\underline d}.{\underline \epsilon}}{\hbar}
\label{dipolerabi}
\ee
where ${\underline \epsilon}$ is the cavity (radiation)  mode 
polarization, and 
\be
E_{vac} \sim  \left(\frac{2\pi \hbar \omega_c}{\varepsilon _0 V}\right)^{1/2} 
\label{amplitude}
\ee
is the r.m.s. vacuum field amplitude at the center 
of the cavity of volume $V$, and of frequency $\omega_c$,
with $\varepsilon _0$ the 
dielectric constant of the vacuum~\footnote{For cavities 
containing other dielectric media, e.g. water in the case of the MT,
$\varepsilon _0$ should be replaced by the dielectric constant 
$\varepsilon$ of the medium.}: $\varepsilon_0 c^2 =\frac{10^7}{4\pi}$,
in M.K.S. units. 
As mentioned above, 
there are simple experiments which confirmed this 
effect~(Bernardot {\it et al.}),
involving beams of Rydberg atoms resonantly 
coupled to superconducting cavities.

The situation which is of interest
to us
involves atoms that are {\it near resonance} with the cavity.
In this case $\Delta << \omega_0$, but such that 
$\lambda^2 N/|\Delta |^2 << 1$; in such a case, formula 
(\ref{rabiabs}) yields two peaks that are characterized
by dispersive frequency shifts $\propto \frac{1}{\Delta}$:
\be
    \Omega \simeq \omega_0 \pm  \frac{N\lambda ^2}{|\Delta|} 
+ {\cal O}(\Delta)
\label{dispersive}
\ee
whilst no energy exchange takes place between atom and cavity mode. 
This is also the case of interest in experiments 
using such Rabi couplings to construct  
Schr\"odinger's cats in the laboratory, 
i.e. macro(meso)scopic combinations
`measuring apparatus + atoms' to verify decoherence 
experimentally. The first experiment of this sort,
which confirms theoretical expectations, is described in ref.~(Brune {\it 
et al.}).

For our purposes in this work, 
it is instructive to review 
briefly the experiment of [Brune {\it et al.} ], where decoherence
of an open mesoscopic system (`Schr\"odinger's cat')
{\it has been demonstrated}. 

\subsection{Rabi splitting and decoherence: experimenal verification}

The experiment consists of sending a Rubidium atom
with two circular Rydberg states $e$ and $g$,
through a microwave cavity storing a small coherent field
$|\alpha>$. The coherent cavity mode is mesoscopic in the sense of
possessing 
an average number of photons of order ${\cal O}(10)$. 
The atom-cavity coupling is measured by the Rabi frequency
$2\lambda /2\pi=48~kHz$. The condition for Rabi dispersive
shifts (\ref{dispersive}) is satisfied 
by having a $\Delta/2\pi \in [70 , 800 ]~kHz$.

The set up of the experiment is as follows:  
The atom is prepared in the superposition of $e,g$ states, 
by virtue of a resonant microwave cavity $R_1$. 
Then it crosses the cavity $C$, which is coupled to a 
reservoir that damps its energy ({\it dissipation})
at a characteristic time scale $T_r << 1.5. ms$. 
Typical cavities, used for atomic scale experiments
such as the above, have 
dimensions which lie in the $mm$ range or at most 
$cm$ range.
In the cavity $C$ a number of photons varying 
from $0$ to $10$ is injected by a pulse source. 
The field in the cavity relaxes to vacuum - thereby 
causing {\it dissipation}
through leakeage of photons through the cavity -
during a time $T_r$, before being regenerated for the 
next atom. The experiment is at an effective 
temperature of $T=0.6K$, which is low enough so as to minimize
thermal effects. 
After leaving $C$, the atom passes through a second
cavity $R_2$, identical to $R_1$; one then measures the 
probability of finding the atom in the state, say, $g$. 
This would mean decoherence. 
The decoherence time is measured for various 
photon numbers; this helps testing the 
theoretical predictions that decoherence
between two `pointer states' of a quantum superposition
occurs at a rate proportional to the square of the distance
among the states~(Gorini {\it et al.} 1978,
Walls and Milburn 1985, Zurek 1991, Ellis {\it at al.} 1989). 

Let us understand this latter point better. 
The coherent oscillator states, characterizing 
the cavity modes, constitute such a pointer basis:
an oscillator in a coherent state is 
defined by the average number of oscillator quanta $n$
as: $|\alpha >:~|\alpha|=\sqrt{n}$. 
Then, consider the measurement of the above-described 
experiment, according to which there is only a phase entanglement
between the cavity and the atom: the combined 
atom-cavity (meter) system is originally in the state
\be
   |\Psi >=|e,\alpha e^{i\phi}>+ |g, \alpha e^{-i\phi}> 
\label{super}
\ee
According to [Brune {\it et al.} 1996] 
the dephasing is atomic-level dependent
$\phi \propto \lambda^2 t n/\Delta$. 
Coupling the oscillator to a reservoir, that damps its energy 
in a characteristic time scale $T_r$, produces decoherence.
According to the general theory~(Ellis {\it at al.} 1984,
Walls and Milburn, Brune {\it et al.} 1991, Gorini {\it at al.} 1978)
the latter 
occurs during a time scale inversely proportional to 
the square of the distance between the `pointer' states $D^2$:
\be
   t_{collapse} = \frac{2T_r}{D^2} 
\label{decoh}
\ee
This is easy to verify in a prototype toy model, 
involving only a phase entanglement, i.e. without 
energy dissipation. To this end, consider the Hamiltonian:
\be
   H=\hbar\omega a^\dagger a + a^\dagger a \Gamma 
\label{phasedamp}
\ee
with $a^\dagger,a$
creation and annihilation operators of a quantum oscillator, and 
$\Gamma $ a phase damping. The `pointer' states
for this problem are characterized by the eigenvalues $n$ 
of the number operator $a^\dagger a$ which commutes
with the Hamiltonian $[ a^\dagger a~,~H]=0$, i.e. 
the pointer basis is $\{ | n > \}$. 
The pertinent Markov master equation
for the density matrix, $\rho $, reads: 
\be
\partial_t \rho =\frac{\kappa}{2}(2a^\dagger a \rho a^\dagger a - 
\rho a^\dagger a a^\dagger a - a^\dagger a a^\dagger a \rho ) 
\label{masterphasedamp}
\ee
Writing $\rho(t)=\sum_{m,n} \rho_{mn}(t)|n><m|$ it is straightforward
to determine the time dependence of $\rho_{nm}(t)$ 
from (\ref{masterphasedamp}) 
\be
   \rho_{nm}(t) = e^{-\kappa (n-m)^2t/2}\rho_{nm}(0)
\label{decoherencedamped}
\ee
which implies that the coherence  
between two different ($m \ne n$) pointer (number) states
is damped at a typical scale given by (\ref{decoh}),
where $T_r = \frac{1}{\kappa}$, and the distance $D$ 
is given by $n-m$ in the above example .  
As mentioned previously, this inverse $D^2$-behaviour 
seems to be a {\it generic} feature 
of open (decohering) systems, including 
open string models~(Ellis {\it et al.} 1992, 1993, Mavromatos Nanopoulos 1997).

In the set up of [Brune {\it at al.} 1996] the distance $D$ is 
given by 
\be
   D=2\sqrt{n}{\rm sin}\phi \simeq 2n^{3/2}\frac{\lambda^2 t}{\Delta}   
\label{distance} 
\ee
for Rabi couplings $2\lambda$, such that $\lambda^2 t n << \Delta $. 
For mesoscopic systems $n \sim 10$, $D > 1$, and, hence, decoherence 
occurs over a much shorter time scale than $T_r$; in particular,
for $\Delta/2\pi \sim 70kHz$ the decoherence time 
is $0.24 T_r$. 
This concludes the construction of a Schr\"odinger's cat,
and the associated `measurement process'. 
Notice that the above construction is made in two stages:
first it involves an interaction of the atom with the 
cavity field, which results in a  coherent state of the 
combined `atom-meter', and then {\it dissipation} is induced 
by coupling the cavity (measuring apparatus) to the environment,
which damps its energy, thereby inducing {\it decoherence}
in the `atom+meter' system. The important point to realize is
that the more macroscopic the cavity mode is (i.e. the higher 
the number of oscillator quanta), the shorter the decoherence
time is. This is exactly what was to be expected 
from the general theory.

\section{MicroTubules as Dielectric Cavities and Dissipationless 
Energy Transfer in the Cell} 

\subsection{Microscopic Mechanisms for the formation of 
Coherent States in MT}

Above we have sketched the 
experimental construction of 
a mesoscopic quantum coherent state
(a `Schr\"odinger's cat' (SC)). 
The entanglement of the atom with the coherent 
cavity mode, manifested experimentally by the `vacuum Rabi splitiing',
leads to a quantum-coherent state for the combined atom-cavity 
system (SC), comprising of the 
superposition of the states of the two-level Rydberg atom.
Dissipation induced by the leakage of photons in the cavity 
leads to decoherence of the coherent atom-cavity state,
in a time scale given by (\ref{decoh}).  
This time scale depends crucially on the nature of the 
coupled system, and the nature of the `environment'. 
It is the point of this session to attempt to discuss a similar 
situation that conjecturally occurs 
in the systems of MT. We believe that understanding the formation of
Schr\'odinger's cats in MT networks will unravel, if true, 
the mysteries
of the brain as a (quantum) computer, which might also be related to
the important issue of `conscious perception', as advocated in 
[Penrose, Nanopoulos, Hameroff and Penrose, Mavromatos and Nanopoulos].

The first issue concerns the nature of the 
`cavity-field modes'. Our point in this section 
is to argue that 
the presence of {\it ordered water},
which seems to occupy the interior of the microtubules (Hameroff 1987),
plays an important r\^ole in producing coherent 
modes, which resemble those of the ordinary electromagnetic 
field in superconducting cavities, discussed above. 

Let us first review briefly some suggestions
about the r\^ole of the electric dipole moment 
of water molecules in producing coherent modes after 
coupling with the electromagnetic radiation field~(Del Giudice 
{\it et al.} 1988). 
Such a coupling implies a `laser-like' behaviour. 
Although it is not clear to us whether such a behaviour characterizes
ordinary water, in fact we believe it does not due to the strong  
suppression of such couplings 
in the disordered ordinary water, however 
it is quite plausible that such a behaviour characterizes 
the ordered water molecules that exist in the interior of MT~(Hameroff 1987). 
If true, then this electric dipole-quantum radiation
coupling will be responsible, according to the analysis of
[Del Giudice {\it at al.} 1988], for the appearance of {\it collective}
quantum coherent modes. 
The Hamiltonian used in the theoretical model of [Del Giudice 
{\it et al.} 1988]
is:
\be
   H_{ow} = \sum_{j=1}^{M} [\frac{1}{2I} L_j^2 + {\underline 
A}.{\underline d}_{ej}]
\label{orwater}
\ee
where $A$ is the quantized
electromagnetic field in the radiation gauge,
$M$ is the number of water molecules, $L_j$ is the
total angular momentum
vector of a single molecule, 
$I$ is the associated (average) moment of inertia,
and $d_{ej}$ is the  
electric dipole vector
of a single molecule, $|d_{ej}| \sim 2e \otimes d_e$,
with $d_e \sim 0.2$ Angstr\"om. 
As a result of the dipole-radiation interaction in 
(\ref{orwater}) coherent modes emerge, which in [Del Giudice 
{\it et al.} 1985, 1986]
have been interpreted as arising from the quantization of the
Goldstone modes responsible for the {\it spontaneous breaking}
of the electric dipole (rotational) symmetry. Such modes are termed
`dipole quanta'. 

This kind of mechanism has been applied to microtubules (Jibu {\it et al.}),
with the conclusion that such coherent modes cause 
`super-radiance', i.e. create a specific quantum-mechanical 
ordering in the water molecules with characteristic times much shorter
than those of thermal interaction. In addition, the optical medium inside the 
internal hollow core of the microtubule
is made transparent by the coherent photons themselves.
Such  
phenomena, if observed, could verify the coherent-mode emission 
from living matter, thereby verifying Fr\"ohlich's ideas.

In our picture of viewing the MT arrangements as cavities,
these coherent modes 
are the quantum coherent `oscillator' modes of section II, 
represented by annihilation and creation operators 
$a_c,a^\dagger_c$, which play the r\^ole of the cavity modes, if 
the ordered-water interior of the MT is viewed as an isolated 
cavity~\footnote{This was not the picture envisaged in
[Del Giudice {\it et al.} 1988]..
However, S. Hameroff, as early as 1974, had conjectured the 
r\^ole of MT as `dielectric waveguides' for photons (Hameroff 1974),
and in [Jibu {\it at al.}] some detailed mathematical construction 
of the emergence of coherent modes out of the ordered water are presented.
In our work in this article 
we consider the implications of such coherent modes
for the system of dimers, in particular for the formation of 
kink solitons of [Satari\'c {\it et. al.}]. Thus, our approach is 
different from that in [Del Giudice 
{\it et al.} 1988, Jibu {\it et al.}], where attention 
has been concentrated only on the properties of the water molecules.
We should emphasize that the phenomenon of optical 
transparency due to super-radiance may co-exist with the 
formation of kink soliton coherent states along the 
dimer chains, relevant to the dissipation-free energy 
transfer along the MT, discussed in the present work.}.
The r\^ole of the small collection of atoms, described in the 
atomic physics analogue above, is played in this picture by the 
protein dimers of the MT chains. The 
latter constitute a two-state system due to 
the $\alpha$ and $\beta$ conformations, defined by the position 
on the unpaired spin in the dimer pockets. 
 The presence of unpaired electrons 
is crucial to such an analogy. 
The interaction of the dipole-quanta coherent modes with the protein dimers
results in an entanglement which we claim is responsible for the 
emergence of {\it soliton quantum coherent states}, extending over 
large scale, e.g. the MT or even the entire MT network.

The issue, we are concerned with here, is whether 
such coherent states 
are 
responsible for {\it energy-loss-free
transport}, as well as for {\it quantum computations}
due to their eventual collapse, 
as a result of `environmental' entanglement
of the entire `MT dimers $+$ ordered water'
system. 
An explicit construction of such solitonic states
has been made in the field-theoretic model for MT dynamics of
Mavromatos and Nanopoulos 1997, 
based on classical ferroelectric models for the displacement 
field $u(x,t)$ discussed in [Satari\'c {\it et. al.}]. 
The quantum-mechanical
picture described here should be viewed as a simplification 
of the field-theoretic formalism, which, however, is sufficient 
for qualitative estimates of the induced decoherence.  

\subsection{Decoherence and dissipationless energy transfer}

To study quantitatively the effects of decoherence in MT 
systems we  
make the plaussible {\it assumption} that 
the environmental entanglement of the `ordered-water cavity' (OWC),
which is responsible for dissipation,
is 
attributed {\it entirely} to the leakage of photons (electromagnetic 
radiation quanta) from the MT interior of volume $V$ (`cavity'). 
This leakage may occur from the {\it nodes} of the MT network,
if one assumes fairly isolated interia. 
This leakage will cause decoherence of the coherent state 
of the `MT dimer-OWC' system.
The leakage determines 
the damping time scale $T_r$ in (\ref{decoh}). 

The dimers with their two conformational states
$\alpha$,$\beta$ play the r\^ole of the
collection of $N$ {\it two-level} Rydberg atoms in the 
atomic physics analogue described in section 3.  
If we now make the assumption that the ordered-water 
dipole-quantum coherent modes 
couple to the dimers of the MT chains in a way similar to the 
one leading to a Rabi splitting, described above, then 
one may assume a coupling $\lambda_0$ of order:
\be
   \lambda _0 \sim \frac{d_{dimer} E_{ow}}{\hbar}
\label{mtcoupling}
\ee
where $d_{dimer}$ is the single-dimer electric dipole matrix element,
associated wiith the transition from the $\alpha$ to the $\beta$
conformation, and $E_{ow}$ is a r.m.s. 
typical value of the amplitude of a coherent dipole-quantum field mode. 

Given that each dimer has a mobile 
charge~(Dustin 1984, Engleborghs 1992): $q=18 \times 2e$, $e$ the electron charge, 
one may {\it estimate} 
\be
d_{dimer} \sim  36 \times \frac{\varepsilon_0}{\varepsilon} \times 
1.6. \times 10^{-19} 
\times 4. 10^{-9} \sim 3 \times 10^{-18}~ {\rm Cb} \times {\rm Angstrom} 
\label{dipoledimer}
\ee
where we 
used the fact that a typical distance for the estimate 
of the electric dipole
moment for the `atomic' transition between the $\alpha,\beta$
conformations is of ${\cal O}(4~{\rm nm})$, i.e. of order of the 
distance between the two hydrophobic dimer pockets. 
We also took account of the fact that, as a result of the 
water environment, the electric charge of the dimers appears
to be screened by the relative 
dielectric constant of the water, $\varepsilon/\varepsilon_0 \sim 80$. 
We note, however, that the biological environment of the unpaired 
electric charges in the dimer may lead to 
further suppression of $d_{dimer}$  (\ref{dipoledimer}). 

The amplitude of the collective modes of the dipole-quanta
may be estimated using the formula
(\ref{amplitude}). In our case  the `cavity volume' is:
\be
V \sim 5 \times 10^{-22}~{\rm m}^3 
\label{volumemt}
\ee
which is a typical MT volume, for a moderately long 
MT $L \sim 10^{-6} {\rm m}$, considered as an isolated 
cavity~\footnote{Here we consider only a single unit of the MT network.
The MT network
consists of a large number of such cavities/MT. Our scenario here 
is to examine soliton formation in a single (isolated) 
cavity MT, with the only 
source of dissipation the leakage of photons. 
This would explain the formation of kinks in a single MT~(Satari\'c 
{\it et al.}, Mavromatos and Nanopoulos).
If the entire network
of MT is viewed as a cavity, 
then the volume $V$ is complicated, but in that case the volume 
can be estimated roughly 
as ${\cal N}_{mt}V$, with $V$ the average 
volume of each MT, and ${\cal N}_{mt}$ the number of MT in the 
network population. In such a case, the solitonic state extends over the 
entire network of MT. It is 
difficult to model
such a situation by simple one-dimensional Hamiltonians as in [Satari\'c 
{\it et al.}, Mavromatos and Nanopoulos].}, 
and $\omega_c$ a typical frequency of the dipole quanta collective mode 
dynamics. To estimate 
this frequency we
assume, following the `super-radiance' model of Jibu {\it et al.},  
that the dominant modes are those 
with frequencies in the range 
$\omega_c \sim \epsilon/\hbar $,
where $\epsilon$ is the energy difference between the two principal 
energy eigenstates of the water molecule, which are assumed
to play the dominant r\^ole in the interaction with the 
(quantized) electromagnetic radiation field. 
For the water molecule:
\be
 \hbar \omega_c \sim 4~{\rm meV}
\label{energy}
\ee  
which yields 
\be
     \omega_c \sim \epsilon/\hbar \sim 6 \times 10^{12} s^{-1} 
\label{frequency}
\ee
This is of the same order 
as
the characteristic frequency of the dimers
(\ref{frequency2}), implying that  
the dominant cavity mode and the dimer system are almost in resonance.
Note that this is a feature shared by  
the Atomic Physics systems in Cavities examined in section 3, and thus 
we may apply the pertinent formalism to our system.

From (\ref{amplitude})
one obtains 
for the r.m.s 
$E_{ow}$ in order of magnitude: 
\be
    E_{ow} \sim 10^{4}~{\rm V/m}
\label{eowmt}
\ee 
where we took into account the relative dielectric constant of water 
$\varepsilon/\varepsilon_0 \sim 80$. 

It has not escaped our attention that 
the electric fields of such order of magnitude 
can be provided by the electromagnetic interactions 
of the MT dimer chains, the latter 
viewed as giant electric dipoles~(Satari\'c {\it et al.}). 
This may be seen to suggest that the super-radiance 
coherent modes $\omega_c$, which in our scenario 
interact with the unpaired electric charges of the dimers 
and produce the kink solitons along the chains, 
owe their existence to  the 
(quantized) electromagnetic interactions 
of the dimers themselves. 

We assume that the system of ${\cal N}$  MT dimers 
interacts with a {\it single} dipole-quantum mode of the ordered 
water and we ignored interactions among the dimer 
spins~\footnote{More complicated situations, including interactions 
among the dimers, as well as of the dimers with more than 
one radiation quanta,
which might undoubtedly occur
in nature, complicate the above estimate.}.  
In our work here we concentrate our attention 
on the formation of a coherent soliton along 
a single dimer chain, the interactions 
of the remaining 12 chains in a protofilament 
MT cylinder being represented 
by appropriate interaction terms in the effective 
potential of the chain MT model of Satari\'c {\it et al.}.  
In a moderately long microtubule of length $L \simeq 10^{-6}~m$ 
there are 
\be
{\cal N} = L/8 \simeq 10^2
\label{nodimers}
\ee
tubulin 
dimers of average length $8~nm$ in each chain.

Then, from 
(\ref{mtcoupling}), 
(\ref{dipoledimer}), and (\ref{eowmt}), 
the conjectured Rabi-vacuum splitting, describing the entanglement
of the interior coherent modes with the dimers can be estimated to be 
of order 
\be 
    {\rm Rabi~coupling~for~MT} \equiv \lambda _{MT} 
= \sqrt{{\cal N}} \lambda_0 \sim 3 \times 10^{11} s^{-1} 
\label{rabiMT}
\ee
which is, on average, an order of magnitude 
smaller than the characteristic frequency 
of the dimers (\ref{frequency2}). 
In this way, the perturbative 
analysis of section 3, for small  Rabi splittings $\lambda << \omega_0$, 
is valid. Indeed, as can be seen from (\ref{frequency2}),(\ref{frequency}),
the detuning $\Delta$ is of order:
\be 
       \Delta / \lambda_0 \sim {\cal O}(10)-{\cal O}(100) 
\label{detun}
\ee
implying that the condition $\lambda_0^2 {\cal N}/|\Delta|^2 << 1$,
necessary for the perturbative analysis of section 3, is satisfied. 
The Rabi frequency (\ref{rabiMT})  
corresponds to an energy splitting $\hbar \lambda_{MT} 
\sim 0.1~{\rm meV}$ for a moderately long MT.

Having estimated the Rabi coupling between the dimers and the ordered-water
coherent modes
we now proceed to 
estimate the average time scale necessary for the 
formation 
of the pointer coherent states which arise 
due to this entanglement~(Zurek 1991, Mavromatos and Nanopoulos 1997). 
This is the same as the decoherence time 
due to the water-dimer coupling.  
In such a case one 
may use the Liouville model for MT of Mavromatos and Nanopoulos, 
discussed in section 2.2, which is known to possess
pointer coherent states~(Ellis {\it et al.} 1994).
The relevant decoherence time scale has been estimated there,
using conformal-field-theory techniques for the dimer chains.  

For an accurate estimate of the decoherence time
one should have a precise knowledge
of the 
violations of conformal invariance associated with the 
water-dimer coupling. Unlike the quantum-gravity induced decoherence
case, where a microscopic model for the description of the 
distortions caused in the environment of the chain 
by the external stimuli is available~(Mavromatos and Nanopoulos), 
at present a precise conformal-field-theory 
model for the description of the water-dimer coupling 
is 
lacking.

However, for our purposes here, 
it suffices to use our 
generic approach to the string-theory
representation of MT chains, described in 
[Mavromatos and Nanopoulos], where an order-of-magnitude estimate for
such violations of conforal invariance 
have been provided, and is argued to be 
{\it maximally violated} of order ${\cal O}(E/M_s)$,
with $E$ a typical low-energy scale, and $M_s$ a `string' scale,
i.e. a typical scale acting as an ultraviolet (upper) cut-off
in energy scales, withi the context of the quantum-integrable model
used for the description of the MT dimer-chains.  

To estimate the `string' scale $M_s$ in our case
~\footnote{In 
[Mavromatos and Nanopoulos], where scenaria for gravity-induced 
conscious perception were discussed, the scale 
$M_s$ was taken to be the Planck scale $10^{19}$ GeV.
In that case 
decoherence was induced by 
the coupling of the entire network 
of brain MT  to quantum-gravity fluctuations
due to the distortion of the surrounding space time, as result
of abrupt conformational changes in the dimers, caused by external
stimuli~(Penrose, Nanopoulos, Hameroff and Penrose, Mavromatos and 
Nanopoulos).},
where only quantum-electromagnetic interactions enter,  
we notice that in the standard string theory 
$M_s$ acts as an ultraviolet cut-off in the energy
of the effective target-sace field theory, in our case the 
field theory of the 
displacement field $u(x,t)$. 
Since in the ferroelectric chain the closest distance 
separating two unpaired electrons, which are assumed to have the dominant
interactions with the water,
is the size of a tubulin dimer 
$\sim d_{min} = 4~{\rm nm}$,  
an order of magnitude estimate of $M_s$ is:
\be 
M_s \sim \hbar v_0/d_{min} \sim 
1.5 \times 10^{-4} {\rm eV}, 
\label{Ms}
\ee
where we took into account that 
the Liouville model for MT in [Mavromatos and Nanopoulos]
is formally  a relativistic string model, but 
with the r\^ole of the velocity of 
light played by the sound velocity $v_0=1 {\rm Km}/{\rm sec}$
in organic biological materials.
It is interesting 
to note that this cut-off in energies 
is less than the kinetic
energies of fast kinks,
propagating with the velocity of sound. Indeed such energies are 
of order $E_{0} \sim 10^{-2} {\rm eV}$. 
Such fast kinks have been associated in Mavromatos and Nanopoulos
with quantum gravitational effects, due to abrupt distortions
of space time, 
and hence their exclusion 
in the approach of the present work, by considering kinks
with energies much less than those, as implied by $M_s$, 
is consistent with our 
considering only electromagnetic interactions
among the dimers and its environment.

The above estimate allows us to 
get an idea of the typical energy scales 
of the excitations of the MT systems
that dominate the ordered-water-dimer 
coupling, in the above scenario, and lead to 
decoherence. 
As discussed in Mavromatos and Nanopoulos
the dominant part of the energy of a 
kink in the model of [Satari\'c {\it et al.}] 
is of order $1 {\rm eV}$, and thus much higher than 
the $M_s$ (\ref{Ms}). 
Such scales may play a r\^ole 
in the decoherence due to quantum-gravity entanglement,
which however is much weaker than the electromagnetic ones considered 
here. On the other hand, as we shall argue now, 
a typical energy scale for the dimer displacement field $u(x,t)$, 
much smaller than $M_s$, and therefore 
consistent with our low-energy approach, 
is provided by the kinetic energy of the kink, which is estimated to be 
of order~(Satari\'c {\it et al.} )
\be
E_{kin} \simeq 5 \times 10^{-8} {\rm eV}
\label{ekin}
\ee
As we shall argue below, 
this energy scale may be used for our estimate of the 
time scale 
necessary for the formation of the coherent state
due to the water-dimer coupling. 
We stresss that (\ref{ekin}) should not be considered as 
a typical energy scale associated with the excitation 
spectrum of the dimers, pertaining to the Rabi splitting
discussed above. 
It is rather an `effective scale', characterising  `friction' effects 
between the dimers and the water environment. The latter are
responsible (Mavromatos and Nanopoulos 1997) for 
decoherence and the eventual formation 
of the kink coherent states, which are 
minimum-entropy states~(Zurek 1991),
least susceptible to the effects of the water environent. 

To justify the above estimate, one should notice that 
the formation of coherent quantum 
states through decoherence due to friction 
is the quantum analogue of 
the `drift velocity' acquired by a Brownian particle 
in classical mechanics.
In the (non-relativistic) conformal-field-theory setting of 
[Mavromatos and Nanopoulos]
such a friction could be described by the formation of point-like defects
on the dimer chains, which could be described by appropriate non-relativistic 
membrane backgrounds in the $1 + 1$-dimensional string theory representation 
of the MT dynamics. Such membranes are stringy defects, which, 
for instance, could 
describe the result of 
an abrupt conformational change of a given dimer due to its coupling 
with the water environment. Scattering of the 
excitations $u(x,t)$
off the defect causes `recoil' of the latter, which starts moving with a 
velocity $v_d$. The recoil is due to quantum fluctuations as argued in 
[Ellis {\it et al.} 1996, 1997].

In [Mavromatos and Nanopoulos]
a lower bound on the time of 
decoherence, due to the presence of such defects, 
was estimated using 
conformal-field-theory methods:
\be 
t_{owdecoh} \gsim   [{\cal O}(v_d^2/16\pi g_sM_s)]^{-1}
\sim 10^{-10}  {\rm sec} 
\label{Mhz}
\ee
for a moderately long MT, with ${\cal N}=10^2$ dimers.
In the above formula, 
$v_d^2$ is the recoil velocity of the defect. By energy-momentum conservation, 
the kinetic energy of the recoiling (non-relativistic) defect
is {\it at most} of the same order as the kinetic energy of the 
displacement field (\ref{ekin}).

This is the time scale over which 
solitonic coherent pointer states in the MT 
dimer system are formed (`pumped'), according to our scenario. 
Note that the 
scale (\ref{Mhz}) 
is not far from the original Fr\"ohlich scale, 
$10^{-11}-10^{-12}~{\rm sec}$ (Fr\"ohlich 1986).

To  
answer the question whether quantum coherent pointer states 
are responsible for loss-free energy transport 
across the MT
one should 
examine the time scale of the decoherence induced by 
the coupling of the MT to their biological environment 
as a consequence of 
dissipation through the walls of the MT cylinders.
Such an ordinary environmental entanglement
has been ignored in the derivation of (\ref{Mhz}).
It is this environment that will induce decoherence 
and eventual collapse of the pointer states formed
by the interaction of the dimers with the coherent modes
in the ordered water. 
Using typical numbers of MT networks, we can estimate
this decoherence time 
in a way similar to the corresponding
situation in atomic physics (\ref{decoh},\ref{distance}): 
\be
    t_{collapse} = \frac{T_r}{2 n {\cal N}{\rm sin}^2\left(\frac{{\cal N}
n \lambda_0^2t}{\Delta}\right)} 
\label{convtime}
\ee
where we took into account 
that 
the dominant (dimer)-(dipole quanta) coupling 
occurs for ordered-water `cavity' modes which are {\it almost at resonance}
with the dimer oscillators (c.f. (\ref{frequency}),(\ref{frequency2})),
slightly detuned by $\Delta:~\lambda_0/\Delta 
<<1 $, c.f. (\ref{detun}).

We also assume that a typical coherent mode of dipole/quanta
contains an average of $n = {\cal O}(1)-{\cal O}(10)$  
oscillator quanta. 
The {\it macroscopic} character of the Schr\"odinger's cat
dimer-dipole-quanta system comes from the ${\cal N}$ dimers 
in a MT (or ${\cal N}_{mt}{\cal N}$ in MT networks). 

The time $t$ appearing in (\ref{convtime})
represents the `time' of interaction of the dimer system with the 
dipole quanta. A reasonable estimate of this time scale 
in our MT case can be obtained by 
equating it with the average 
life-time of 
a coherent dipole-quantum state, which, in the 
super-radiance model of [Jibu {\it et al.}]
can be estimated as
\be
     t \sim \frac{c\hbar ^2V}{4\pi d_{ej}^2\epsilon N_{w}L}
\label{lifetime}
\ee
with $d_{ej}$ the electric dipole moment of a water molecule, 
$L$ the length of the MT, and $N_w$ the number of water molecules 
in the volume $V$ of the MT. For typical values of the parameters
for moderately long MT, $L \sim 10^{-6}~{\rm m}$, $N_w \sim 10^8$, 
a typical 
value of $t$ is: 
\be 
         t \sim 10^{-4}~{\rm sec} 
\label{lifetime2}
\ee

We remark at this point that this 
is considerably larger than 
the average life-time of a coherent dipole quantum 
state 
in the water model 
of [Del Giudice {\it et al.} 1988]. Indeed in that model, 
the corresponding life-time is estimated to be:
\be 
     t \sim 2\pi/\omega_0 
\label{waterdecoh}
\ee
where $\omega_0 \sim 1/I$, is a typical frequency 
of resonating electromagnetic mode in the ordered water.
For a typical value of the water molecule moment 
of intertia~(Del Giudice {\it et al.})
this yields a time-scale associated with the coherent
interaction ${\cal O}(10^{-14}~{\rm sec})$.

The time scale $T_r$, over which a cavity MT dissipates its energy,  
can be identified in our model with the average life-time 
(\ref{lifetime}) of a coherent-dipole quantum state: 
\be
T_r \sim t \sim  10^{-4}~{\rm sec}
\label{trfrohlich}
\ee
which leads to a naive estimate 
of the quality factor for the MT cavities,
$Q_{MT} \sim \omega_c T_r \sim {\cal O}(10^8)$. 
We note, for comparison, that high-quality 
cavities encountered in Rydberg atom
experiments 
dissipate energy in time scales of ${\cal O}(10^{-3})-{\cal O}(10^{-4})$
sec, and have $Q$'s which are comparable to $Q_{MT}$ above. 
Thus, it is not unreasonable to expect that conditions 
like (\ref{trfrohlich}),
characterizing MT cavities, are met in Nature.

{}From (\ref{convtime}), (\ref{rabiMT}), and (\ref{trfrohlich}), 
one then obtains the following estimate 
for the collapse time of the kink coherent state of the MT dimers 
due to dissipation:
\be
t_{collapse} \sim {\cal O}(10^{-7})-{\cal O}(10^{-6})~{\rm sec}
\label{tdecohsoliton}
\ee
which is larger or equal than the scale (\ref{FS})
required for energy transport 
across the MT by an average kink soliton in the model of 
[Satari\'c {\it et al.}]. The result (\ref{tdecohsoliton}), then, 
implies that 
Quantum Physics may be responsible for 
dissipationless energy transfer across the MT.

\begin{figure}
\begin{center} 
\begin{picture}(320,320)(70,0)
\SetPFont{Helvetica}{10}
\SetScale{1}
\DashLine(200,200)(200,50){5}
\DashLine(300,200)(300,50){5}
\Line(150,200)(150,50)
\Line(149,200)(149,50)
\Line(148,200)(148,50)
\Line(350,200)(350,50)
\Line(351,200)(351,50)
\Line(352,200)(352,50)
\Oval(250,200)(40,100)(0)
\Oval(250,50)(40,100)(0)
\Oval(250,50)(10,50)(0)
\Oval(250,200)(10,50)(0)
\end{picture}
\end{center}
\caption{Schematic Representation of the 
{\it interior} of a MT Arrangement in a cell.
The thick vertical lines represent the interior walls of the MT cylinders, 
consisting of protein dimers 
in $\alpha$ or $\beta $ conformations, depending on the `location' of the unpaired charges. In the quantum mechanical
picture each dimer finds itself in a {\it superposition} of $\alpha$ and 
$\beta$ states.
The interior of the MT cylinder 
is assumed to be full of `ordered water' molecules. 
The {\it thin interior layer}, of thickness 
of a few Angstr\"oms, located in the region 
between the dimer walls and the 
dashed lines, indicates probable 
regions that are thermally isolated, and can operate
as electromagnetic cavities, thus sustaining the dimer 
coherent states for some time.}
\label{fig1}
\end{figure}
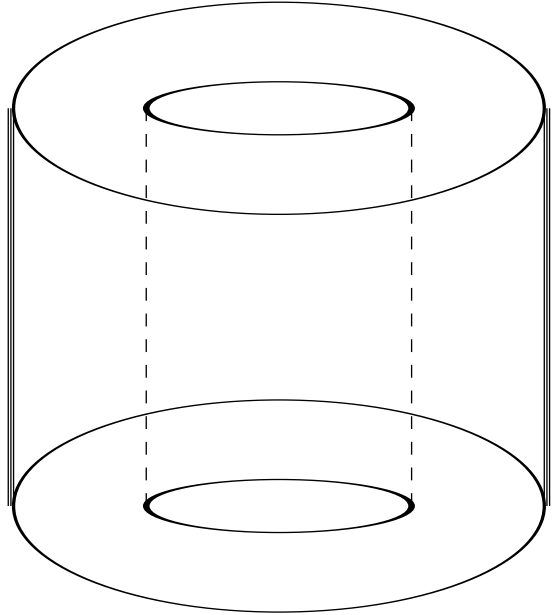

Before concluding we would like to stress that above 
we have presented the most optimistic scenario, according to which 
the whole interior of the microtubule cylinders acts as a
(thermally) isolated quantum electromagnetic cavity. 
However, such a spatial volume,
of cross-section diameter 
$16~nm$,  
might be too big for this to be true.
What we envisage as happening in realistic situations is 
the existence of a {\it thin layer} near the dimer walls, of thickness
of up to a few Angstr\"oms, 
which acts as a cavity volume in which the electric 
dipole-dipole interactions between water molecules and dimers,
dominate 
over the thermal losses (see figure \ref{fig1}).

Indeed, 
if we consider two electric dipole vectors 
$\underline{d}_i$, $\underline{d}_j$, 
at locations $i$ and $j$
at a relative distance $r_{ij}$, 
one pertaining to a water molecule, and the other to a protein dimer
in the MT chain, 
then the dipole-dipole interaction has the form:
\be
E_{dd} \sim -\frac{1}{4\pi \epsilon}\frac{3 ({\hat \eta}.{\underline d_i})
({\hat \eta}.{\underline d_j})- \underline{d}_i . \underline{d}_j}{|r_{ij}|^3}
\label{dipoledipole}
\ee
where ${\hat \eta}$ is a unit vector in the direction of $\underline{r}_{ij}$,
and
$\epsilon $ is the dielectric constant of the medium. 
Taking the medium 
between the ordered-water molecules in the layer 
and the MT dimers
to correspond to the tubulin protein, and using 
a typical value $\epsilon \sim 10$ (Tuszynski and Brown 1997),
as well as the generic (conjectural) 
values of the electric dipoles for tubulin dimers 
and water molecules given above, it is easy to see that 
the dipole-dipole interactions (\ref{dipoledipole}) 
may become of the same order, and/or overcome thermal losses 
at room temperatures, $\sim k_BT$, 
for $|r_{ij}|$ of up to a few tenths of an Angstr\"om. 
Notice that for such distances the respective order of the 
energies is ${\cal O}(10^{-2}~{\rm eV})$. 
Such thin cavities may not be sufficient, however, to sustain 
quantum coherent modes for sufficiently long times so that energy 
transfer is acomplished along the MT. 

At this point we cannot resist in remarking that 
isolation from thermal losses could be assisted enormously by 
the existence of a {\it ferroelectric} transition below some critical 
temperature $T_C$, for the system of protein dimers in MT. For 
our scenario, discussed above, it would be ideal if such a temperature were 
close to 
room temperatures. It is remarkable that such ferroelectric 
transitions, occuring around $T_C \sim 300$ K, 
are encountered in  lattice models of MT dynamics 
existing in the literature (Tuszynski {\it et al.} 1995).

To understand better the r\^ole of ferroelectricity in 
the context of our quantum-coherent state situation, it 
useful to view the MT dimers as 
a 
lattice system, resembling `ion lattices' in real solids.  
As is well known, in the ion lattice case 
(Ziman 1972) a ferroelectric transition 
induces a dynamic dielectric `constant' $\epsilon (\omega )$
for a moving electron 
of frequency $\omega$:
\be
   \epsilon (\omega ) \simeq \epsilon (\infty) + 
\frac{\Omega _p^2}{\omega _T^2 - \omega ^2}
\label{dielectric}
\ee
where $\Omega _p^2$ is some constant that depends on the 
ion lattice, 
$\epsilon (\infty)$ represents the dielectric constant of the medium
measured for $\omega >> \omega _T$, and
$\omega _T^2$ represents `ion lattice' vibrations, depending on 
temperature. Temperature effects introduce imaginary parts in the 
denominator of the right-hand-side of (\ref{dielectric}), 
and in the ferro-electric case such effects are capable of inducing - 
for a specific set of `lattice displacements' - negative $\omega _T^2$
(Ziman 1972). In such a case, for certain frequencies 
\be
  \omega ^{*2} =\frac{\Omega _p^2}{\epsilon (\infty )}-|\omega _T^2|
\label{ferroel}
\ee
the medium may look as if it is characterized by 
a {\it vanishing} (dynamical) dielectric `constant', 
which has an obvious effect
in inducing large electric field variations 
at this frequency, without external excitations. 
In addition, for a frequency range $\omega ^2 < \omega ^{*2}$ 
the dielectric tensor may become {\it negative}, which would imply that 
the medium becomes {\it opaque} for this range of frequencies.  
The advantage of ferroelectricity, as compared to the 
conventional media, is that the critical frequency $\omega^*$ 
is considerably smaller than the one 
in non-ferroelectric cases 
(where $\omega _T ^2 >0$).

In our scenario of representing the MT as cavities,
the r\^ole of the moving electron could be played by 
the {\it dipole quanta} (ordered-water coherent modes)~\footnote{After all,
the electric dipole moment of the water molecules is due to a displacement 
of two electrons (Del Giudice {\it et al.} 1985, 1986)}, 
which are characterized by frequencies $\omega _c$ (\ref{frequency}). 
One would really like to identify  $\omega ^*=\omega _c$,
in which case the medium in the thin layers of figure \ref{fig1}
would appear to have a vanishing,
or in practice smaller than one, relative  dielectric `constant', 
for certain frequency modes, 
thereby screening thermal losses in regions around the walls
of the MT (see figure \ref{fig1})
which could be substantially 
thicker than in the non ferrroelectric case.
Indeed for effective dielectric constant $\epsilon (\omega) < 1$
(\ref{dipoledipole}) can overcome thermal losses at room 
temperatures for up to {\it a few} Angstr\"oms in realistic situations.   
An additional possibility would be that for this range of frequencies 
a {\it negative} (dynamical) dielectric constant arises. 
This  
would mean 
that the dimer walls become {\it opaque} for the modes
in the range of frequencies $\omega ^2 < \omega ^{*2}$, 
thereby supporting the idea of isolated `cavities' put forward by 
[Mavromatos and Nanopoulos 1997b]. As explained above, such frequencies
are not high, and in fact may occur within the range of 
frequencies discussed in section 2.

Notice that the model of [Satari\'c {\it et al.}], used in our work, 
is based on a ferroelelctric model, where the ferroelelctrism is due to 
the dipole-dipole interactions among the protein dimers. 
We note in passing, that such interactions are usually hold responsible 
in `first-principles scenaria' for the theory of 
ferroelelctrism in materials
(W. Zhong {\it et al.}., 1995, Ph. Ghosez {\it et al.} 1996). 
In fact, in the standard atomic physics
situations,
the ferroelectric instability 
in the lattice, yielding $\omega _T^2 < 0$, is viewed as a 
consequence of a 
delicate balance between the dipole-dipole interactions in the crystal,
and short-range forces in the lattice ions.  
It is our belief that 
the MT arrangements exhibit ferroelectric properties,
at critical temperatures which are close to room temperatures, 
and
we think that a laboratory demonstration of this 
will be a very challenging experiment to 
perform (Zioutas 1997, Mavromatos, Nanopoulos, Samaras and Zioutas 1997).

It can be shown that, in the case where the cavity 
thickness extends up to a few atomic scales,  
the results of the above analysis, 
especially those related to the 
order of magnitude of the super-radiance life time (\ref{lifetime})
and  
the collapse time (\ref{tdecohsoliton}), 
are not affected much.  

We also 
notice that it has been recently confirmed experimentally 
that in the {\it exterior} of the MT cylinders, there are 
{\it thin} layers of charged ions, of thickness of order a ${\cal O}(7-8)$
Angstr\"oms, 
in which the electrostatic 
interaction is larger than the thermal energy due to the interaction with 
the environment (Sackett 1997). 
In view of such result we conjecture that similar layers might 
exist in the interior of the MT cylinders, which provide 
us with the necessary thermal isolation to sustain quantum coherent 
states over time scales of order (\ref{tdecohsoliton}).
Whether this is happening in all cell MT, or only in certain 
areas, such as brain MT, is something that we are agnostic about, at present. 
It should be noticed that questions like these can only be answered when 
precise information, at an atomic scale, becomes available   
on the structure of tubulin dimers, on the magnitude of 
their electric dipole moments, 
and on the detailed structure of the water interia of MT. 
As a first step towards this direction
we mention the atomic resolution map of tubulin,
which became available only very recently 
by means of electron crystallogaphy (Nogales, Wolf and Downing 1998).

We close this section by remarking that 
if the condition $T_{collapse} \gsim 5 \times 10^{-7} {\rm sec}$
is not met, then 
the above picture, based on mesoscopic coherent states, 
would be inconsistent 
with energy-loss free energy transport, since 
decoherence due to environmental entanglement 
would occur before energy could be transported across the 
MT by the preformed quantum soliton.
In such a case energy would be transported 
due to different mechanisms, one of which is 
the 
classical solitons scenario of [Lal, Satari\'c {\it et al.}]. 
However, even in such cases of fast decoherence, the 
Rabi coupling predicted above could still exist and be 
subjected to experimental verification.

\section{MicroTubules  as Quantum Holograms}

In the previous discussion we have argued in favour of the 
existence of {\it coherent quantum modes} in the interior and walls
of the MT arrangements, when viewed as 
cylindrical electromagnetic cavities. 
In this section we would like to conjecture, based on the 
above considerations,  
a holographic scenario for information processing.

Holography is based on the {\it interference} of two 
kinds of waves, a `reference wave' and an `object wave'. 
In standard optical holographic devices, a {\it laser} beam 
of (coherent) light is essentially {\it split} into two beams, 
one to illuminate the subject, and one to act as a reference. 
The hologram is then produced by the {\it interference} 
of these two waves. 

MT have been conjectured to operate, in generic terms, as 
holographic devices for information processing and memory 
printing (Pribram 1991, Jibu {\it et al.}). 
The coherent photons due to the super-radiance
scenaria 
of [Jibu {\it et al.}] and [Del Giudice {\it et al.} 1988]
in the water interior of MT, may be considered as providing the coherent 
source of light required to produce holographic 
information processing. This might be important for information 
processing and memory of brain cells. 

However, 
there is problem with considering the water interior 
regions of MT as being {\it solely} responsible for information processing, 
and memory printing. The problem 
is associated with the fact that 
such processes store information in a {\it code},
which however has too little structure to avoid 
{\it memory overprinting} (Vitiello 1996).  

Indeed, according to the  
{\it quantum scenaria} 
for the function of biological systems, memory - and more generally 
information processing by the brain - may correspond 
simply to a spontaneous breaking of a certain symmetry
by the ground state of the biological system 
(Ricciardi and Umezawa 1967). 
Memory in this picture corresponds to excitation 
by the external stimulus
of the 
Goldstone modes arising from the spontaneous 
breakdown of the symmetry, whose quantum numbers 
constitute a special {\it coding}. 
Memory recall corresponds to the excitation by another external stimulus
of the {\it same} quantum numbers.
The important question in this respect 
is what kind of symmetry is involved in such problems. 

In [Del Giudice {\it et al.} 1985-1988] the symmetry was assumed 
to be the dipole symmetry, which is argued to be sponaneously broken 
in the water interia of MT, producing coherent super-radiant modes. 
However, this symmetry, which is basically an $O(3)$ symmetry, 
produces only 
a limited number of quantum numbers, thereby leading to 
a limited memory capacity, leading inevitably to 
{\it overprinting}:
once the set of quantum numbers 
is exhausted by a succession of stimuli in a memory process,
then, subsequent stimuli will excite the same quantum numbers,
with the inevitable result of erasing previously stored information,
({\it memory overprinting}). 

As a way out of such a limited memory capacity problems, 
Vitiello (Vitiello 1995) 
suggested that the very fact that the 
brain was an {\it open} quantum system, interacting with 
a {\it dissipative} environment, was sufficient to provide
an {\it infinity} of degrees of freedom, thereby increasing 
enromously the capacity of the brain cells to store practically 
unlimited information. However, the suggestion of 
[Vitiello] was rather
generic, without making an attempt to present 
a microscopic model for the brain, where such a scenario could be realized.

In [Mavromatos and Nanopoulos 1997a,b] a more microscopic treatment 
of the brain as an open quantum 
sytem was given, by arguing that the dimer chains 
of the MT in brain cells provide a {\it quantum integrable} 
system of electric dipoles, interacting with other chains.
Such one-dimensional chains of dimers 
are characterised by infinite dimensional symmetries, 
termed $W_\infty$ symmetries (Bakas and Kiritsis 1992).
Such symmetries mix the propagating low-energy modes
of electric-dipole quanta, with non-propagating global 
(topological) degrees of freedom of the MT chain structures,
representing the `environment' of the chain, and providing 
a coding system with an {\it infinite} capacity. 
It should be mentioned that 
such global modes may be thought of as representing high `spin' 
excitations in the `effective two-dimensional' space-times of the dimer 
chains, and are separated by the 
`low-lying' propagating dipole quanta by gaps in the spoectrum,  
of order $>M_s$, the characteristic scale, (\ref{Ms}), 
of the quantum integrable model 
discussed in section 3.

In these scenaria, it becomes clear that if the processes
of memory coding and memory recall are to 
be associated with a kind of holographic processing 
and storage of information, 
then the interference patterns must be created by waves
in the dimer lattice rather, than the bulk of the water interia
of MT. Notice, therefore, 
that the {\it paracrystalline} structure of the MT dimer lattice, 
which was argued previously to be an essential feature for 
quantum coherence and ferroelectricity, it also plays an
important r\^ole within the context of holographic
information processing.

\begin{figure}
\begin{center} 
\begin{picture}(320,320)(70,0)
\SetPFont{Helvetica}{10}
\SetScale{1}
\GCirc(200,200){70}{1}
\GCirc(200,200){60}{1}
\GCirc(200,200){50}{1}
\GCirc(200,200){40}{1}
\GCirc(200,200){30}{1}
\GCirc(200,200){20}{1}
\GCirc(200,200){10}{0}
\ArrowLine(200,200)(400,250)
\GText(100,200){1}{Object Atom} 

\GCirc(200,100){70}{1}
\GCirc(200,100){60}{1}
\GCirc(200,100){50}{1}
\GCirc(200,100){40}{1}
\GCirc(200,100){30}{1}
\GCirc(200,100){20}{1}
\GCirc(200,100){10}{0}
\ArrowLine(200,100)(400,150)
\GText(100,100){1}{Source Atom}
\Line(400,300)(400,50)
\G2Text(450,200){1}{Far Field}{Detector}
\ArrowLine(200,100)(200,200)

\end{picture}
\end{center}
\caption{Schematic Representation of a Holographic Arrangement
for an internal source X-ray Holography.
The Hologram is produced in the far field detector by the interference
of the spherical waves from the source and object atoms.
The arrows indicate direction of wave propagation.}
\label{fig1b}
\end{figure}
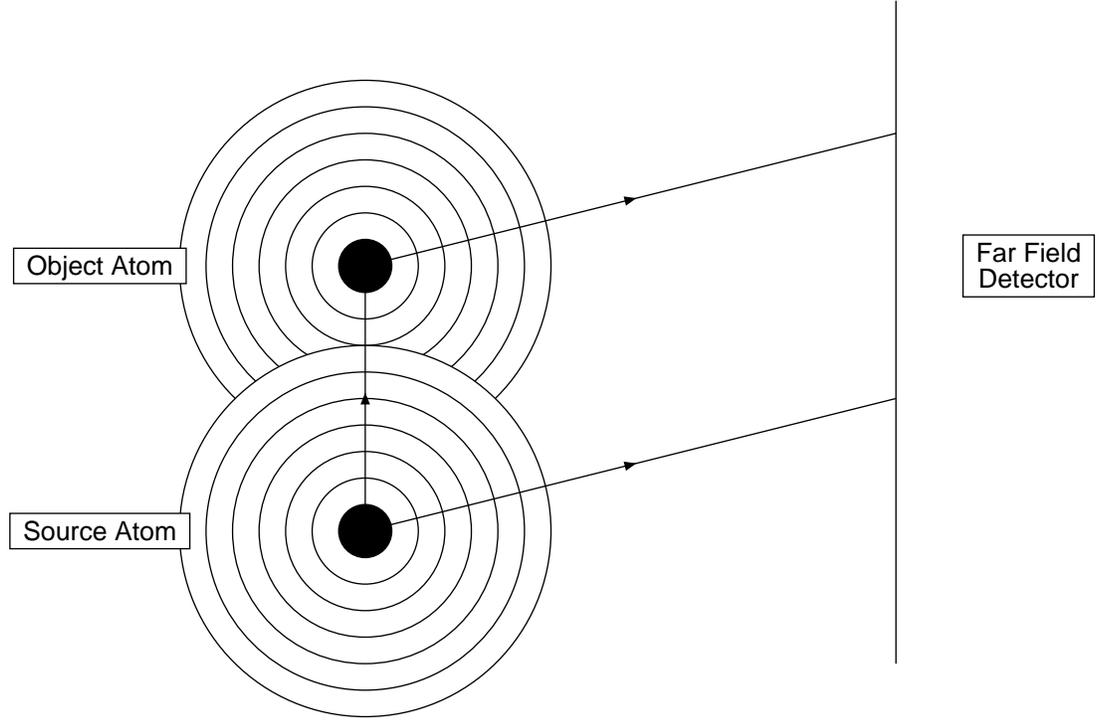

Below we conjecture a situation for the MT dimer 
protofilament chains as holographic information
processors,  
which resembles that of 
internal source X-ray holography in atomic physics 
(Miller and Sorensen 1996). In that type of holography,
there is an internal source of particles, 
inside the atomic crystal, which leave the crystal 
in nearly spherical wave states. This source produces
a strong reference beam. 
These particles
{\it coherently} single-scatter from the object atoms in the 
crystal. The {\it interference} between these single-scattering 
events and the strong direct path of the reference beam 
produces the hologram (see figure \ref{fig1b}).

In [Miller and Sorensen] it was argued that,despite the fact that the 
Holography equations can be derived from simple classical wave physics 
for the description of the interference of the reference and object waves, 
however {\it quantum electrodynamical effects}, and in particular 
{\it virtual photons and electrons }, i.e. intermediate particles -
not on the mass shell - exchanged in quantum corrections, 
play an important r\^ole in  producing these equations.

The key ingredients in the internal source X-ray holography
is the {\it localized} source of the coherent X-ray photons
lying inside the atomic sample, and the {\it coherent} 
scattering event at the object atoms, without which 
the interference final state will not occur.

Such X-ray photons could be 
produced by 
Bremsstrahlung processes (Bjorken and Drell 1964),
according to which an electron $e(p)$ incident on a solid radiates a photon 
$\gamma (k) $ due to the excitation of the solid:
\be
    e(p_i) + {\rm solid} \rightarrow e (p_f)' + \gamma (k) + {\rm excited}~{\rm solid}'
\label{brems}
\ee
The hologram is produced by the {\it quantum mechanical} interference 
of the direct photon amplitude, i.e. the amplitude for the 
photon to leave the solid without interactions, which serves as the reference
wave, and the wave corresponding to the amplitude produced by the single 
photon-atom scattering, which plays the r\^ole of the 
object wave. Multiple scattering events are suppressed for hard X-rays, 
as being weak compared to the strong reference wave, and this is an 
ideal holographic situation. 
The virtual photon and electron 
effects appear in the computation of the respective amplitudes,
and, as argued in [Miller and Sorensen], are important in yielding
the correct holography (interference) equations. 
It is understood that in order  to have a high quality 
hologram, the wavelength of the photons should be smaller than the 
distance between the atoms in the crystal. 

Let us now see how a situation like the above can be realized in 
our scenario of viewing the MT as electromagnetic cavities. 
The r\^ole of the atomic crystal is played by the 
protein dimer cylindrical lattices. The fact that 
in Nature all MT arrangements are characterized by 13 protofilaments, 
probably imply - in conjunction with the holographic picture - some special 
information 
coding, yet to be understood.  
The external stimulus provide the necessary distortion 
on the dimer lattice, so as to cause the emission 
of coherent modes from a `localized source' in the dimer lattice. 
This is the analogue of the Bremsstrahlung process 
(\ref{brems}) in the X-ray holography.  
The r\^ole of the coherent beam of particles produced in the 
localized atomic source is played by the quantum coherent solitonic modes
(dipole quanta) of [Mavromatos and 
Nanopoulos], arising from decoherence effects due to the water environment
of
the dimer chains in the way explained above. 
The coherent nature of the solitonic mode is essential in 
yielding single scattering events between the solitonic mode  
and the 
neighboring dimers and/or water molecules in the cavity region
of MT (see figure \ref{fig1}). The holographic information 
processing in such a picture 
is produced by the
{\it quantum mechanical 
interference} between the direct amplitude for 
these coherent modes, i.e.  
the amplitude for the mode to pass through 
the MT arrangement without interactions, and 
the amplitude for the single scattering of the dipole coherent 
quanta on the neighboring dimers
and water molecules.

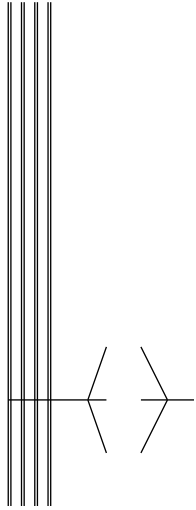
\begin{figure}
\begin{center} 
\begin{picture}(320,320)(70,0)
\SetPFont{Helvetica}{10}
\SetScale{1}
\Line(200,200)(200,10)
\Line(201,200)(201,10)
\Line(206,200)(206,10)
\Line(205,200)(205,10)
\Line(210,200)(210,10)
\Line(211,200)(211,10)
\Line(215,200)(215,10)
\Line(216,200)(216,10)
\Line(200,50)(230,50)
\Line(230,50)(237,70)
\Line(230,50)(237,50)
\Line(230,50)(237,30)
\Line(250,70)(260,50)
\Line(250,50)(260,50)
\Line(250,30)(260,50)
\Line(260,50)(270,50)
\end{picture}
\end{center}
\caption{Schematic Representation of MT Arrangement in a neuron
cell. The MTs are represented as vertical thick lines. 
The synapsis is represented by a discontinuity 
on the thin line that emerges vertically from the MT arrangements,
on the lower right-hand-side part of the figure. Information 
is argued to be recorded in the form of a hologram on the vesicular grid,
which is the part of the synapsis lying left of the discontinuity. 
The grid plays the r\^ole of the far-field detector in
fig. \ref{fig1b}. Quantum Mechanical
phenomena might be important in transmitting 
information through the synapsis via tunneling
(`probabilistic firing' from the vesicular grid).}
\label{fig3}
\end{figure}

If the above holographic scenario is realized also in 
brain cell MT, then it might well be that 
the relative information on the phase and the amplitude of the `object wave', 
contained in the interference
fringes, 
is recorded by the synapsis
(see figure \ref{fig3}), which plays the r\^ole of the far field detector
in X-ray holographic devices (see figure \ref{fig1b}).  
Notice that if this scenario is true,
the resulting hologram is of high quality, since the 
wavelength of the coherent dipole modes on the dimer chains 
is smaller than the distance between the dimers. 
Indeed, adopting as a typical frequency of the mode 
(\ref{frequency2}), and taking into account that the phase 
velocity is $2 m/sec$ (Satari\'c {\it et al.}), one sees immediately that 
the wavelength is of order $10^{-3}~{\rm nm}$, which is much smaller 
than the distance between neighboring dimers ($\sim 4~{\rm nm}$). 
Notice that in this holographic picture, memory processes 
in the brain would correspond to 
a storage of information concerning the quantum numbers of 
the Goldstone modes arosen by the spontaneous breaking, as a result 
of the external stimulus,  
of the 
infinite $W_\infty$ symmetry  
characterising global excitations 
modes of the MT chains, viewed as `dissipative  quantum systems' 
(Mavromatos and Nanopoulos 1997a).

The recording of information in the synapsis takes place in the
vesicular grid of the presynaptic region (Eccles 1986): the vesicles 
of the grid receive the interference patterns from the holographic
process, and in this way the information about the 
distortion due to the external stimulus is recorded.
Notice that due to the above-mentioned high resolution hologram, 
quite detailed information is received about the position 
of the structures in the neighborhood of the dimer-`source'
(position of dipoles of dimers and/or  water molecules, global excitations
of the dimer chains etc). This information 
is stored in the presynaptic grid like a quantum hologram,
which is then used in the memory recall process, as explained above.
In this way, the presynaptic vesicular grid may be viewed as
playing the r\^ole of the far-field detector in fig. \ref{fig1b}. 

Information is then processed across the synapsis by 
{\it quantum tunnelling} via the probabilistic `firing' 
of the vesicles in the paracrystalline grid of the presynaptic 
region (Eccles 1986, Beck and Eccles 1992).
Notice that the life time $10^{-7}~{\rm sec}$ of the dimer coherent 
mode, that we have argued in this work to be sufficient for 
energy-loss free transport across the MT, is also sufficient for 
the above-described process of holographic 
information processing.

It should be stressed once more 
that, in the context of the microscopic 
quantum integrable model of [Mavromatos and Nanopoulos 1997a], 
in addition to the storage of information pertaining to the propagating 
coherent solitonic modes of the dimer-dipole quanta, there is a 
`generalized' holographic processing and storage 
of information pertaining to the $W_\infty$ discrete, delocalised,  
{\it non-propagating} global modes, 
referring to higher `spin excitations' 
in the `effective space-time' of the dimer chains. 
The generalized holographic picture is due to the 
non-trivial interactions 
among such global modes (Mavromatos and Nanopoulos 1997a), despite  
their non-propagating, topological, nature. The excitation of the 
corresponding quantum numbers was argued to be responsible for 
an enormous enhancement of the memory capacity, as mentioned previously. 

It is understood that at the present stage
the above considerations should be considered as 
conjectural. Unfortunately, due to the complexity of the 
tubulin dimer structure, and the lack of knoweldge 
of the respective map at an atomic resolution level,
it is not yet possible to construct 
microscopic models that could possible allow one to derive the holographic 
equations for the coherent waves from a 
semi-microscopic quantum field theory model,
as happens in the internal source X-ray holography
(Miller and Sorensen). For such a purpose, one might have to 
wait 
for a complete understanidng  of the structure 
of the tubulin protein at an  atomic resolution level 
(Nogales {\it et al.}). Only then, it may be possible to 
perform (numerical) studies of possible models 
that could test the above-conjectured holographic information 
processing.

\section{Outlook}

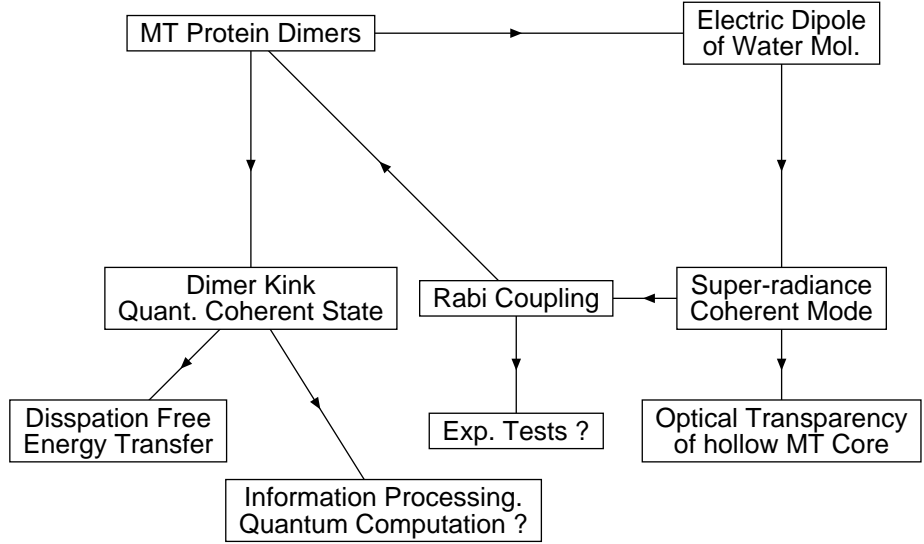
\begin{figure}
\begin{center} 
\begin{picture}(320,320)(0,0)
\SetPFont{Helvetica}{10}
\SetScale{1}

\ArrowLine(100,200)(100,100)
\ArrowLine(100,200)(300,200)
\ArrowLine(300,200)(300,100)
\ArrowLine(300,100)(300,50)
\ArrowLine(300,100)(200,100)
\ArrowLine(200,100)(100,200)
\ArrowLine(200,100)(200,50)
\ArrowLine(100,100)(50,50)
\ArrowLine(100,100)(150,20)

\GText(100,200){1}{MT Protein Dimers}
\G2Text(100,100){1}{Dimer Kink}{Quant. Coherent State}

\G2Text(300,200){1}{Electric Dipole}{of Water Mol.}

\G2Text(300,100){1}{Super-radiance} 
{Coherent Mode}

\G2Text(300,50){1}{Optical Transparency}{of hollow MT Core}

\GText(200,100){1}{Rabi Coupling}
\GText(200,50){1}{Exp. Tests ?} 

\G2Text(50,50){1}{Disspation Free}{Energy Transfer}

\G2Text(150,20){1}{Information Processing.}{Quantum Computation ?}
\end{picture}
\end{center}
\caption{Suggested `Basic Cycle' for dissipationless
energy transfer in cell MT due to Quantum Physics.
The dominant interactions between dimers 
and water molecules are asumed to be electromagnetic.}
\label{fig2}
\end{figure}

In the present work we have put forward a conjecture
concerning the representation of the MT arrangements 
inside the cell as {\it isolated} high-Q(uality) {\it cavities}. 
We presented a scenario according to which 
the presence of the ordered water in the interior of the cylindrical 
arrangements results in the appearance of electric dipole quantum 
coherent modes, which couple to the unpaired electrons of the 
MT dimers via Rabi vacuum field couplings, familiar from 
the physics of Rydberg atoms in electromagnetic cavities
(Sanchez-Mondragon {\it et al.}). 
In quantum optics, such couplings are considered as experimental 
proof of the quantized nature of the electromagnetic radiation.
In our case, therefore, if present, such couplings could 
indicate the existence of the coherent quantum modes
of electric dipole quanta in the ordered water environment of MT,
conjectured in [Del Giudice {\it et al.} 1985, 1986, 1988], 
and used in the present work.
Our mechanism for the emergence of coherent states 
in MT arrangements inside the cell can be summarized
by the `basic cycle' shown in figure \ref{fig2}. 

Some generic decoherence time estimates, due to environmental 
entanglement of the MT cavities, have been given. The conclusion is 
that only in {\it fairly isolated} cavities, which we conjecture exist
inside the biological cells, 
decoherence occurs in time scales which are in agreement 
with the Fr\"ohlich scale ($\sim 5 \times 10^{-7}$ sec) for 
energy transfer across the MT via the formation 
of kink solitonic structures. In such a case,  
{\it dissipationless energy transfer} might occur in biological systems,
an in particular in MT which are the substratum of cells, 
in much the same way as frictionless electric current transport occurs
in superconductors, i.e. via {\it quantum coherent modes} that extend over 
relatively large spatial regions. A phenomenological analysis 
indicated that for moderately long MT networks such a situation could be 
met if the MT cavities dissipate energy in time scales 
of order $T_r \sim 10^{-4}-10^{-5} {\rm sec}$. This lower bound 
is comparable to the 
corresponding scales of atomic cavities~(Bernardot {\it et al.} 1992,
Brune {\it et al.} 1996), 
$\sim 10^{-4} {\rm sec}$, which, in turn, implies that the 
above scenaria, on dissipationless energy transport 
as a result of the formation of quantum coherent states,
have a good chance of being realized at the scales 
of MT, which are comparable to those in Atomic Physics.

We have conjectured in this work
that an indirect verification of such a mechanism would be the 
experimental detection of the aforementioned Vacuum field Rabi coupling,
$\lambda_{MT}$,  
between the MT dimers and the ordered water quantum coherent modes. 
This coupling, if present,  
could be tested 
experimentally  
by the same methods used to measure VFRS in atomic physics~(Bernardot 
{\it et al.}), 
i.e. by using the MT themsleves as {\it cavity environments}, 
and considering tuneable probes to excite the coupled dimer-water
system. Such probes could be pulses of (monochromatic) light, for example,
passing through the hollow cylinders of the MT. 
This would be the analogue of an external field in the atomic experiments
described above, which would then resonate, not at the bare frequencies 
of the coherent dipole quanta or dimers, but at the {\it Rabi splitted} ones,
and this would have been exhibited by a double pick in the absorption 
spectra of the dimers. By using MT of different sizes
one could thus check on 
the characteristic $\sqrt{N}$-enhancement of the Rabi coupling 
for MT systems with $N$ dimers~\footnote{The technical complications 
that might arise in such experiments are associated 
with the absence of completely
resonant cavities in practice. In fact, from our discussion in this article,  
one should 
expect a slight detuning $\Delta$ between the cavity mode and 
the dimer, of  
frequency $\omega _0$. As discussed in section 3,
the detuning produces a split of the vacuum-Rabi 
doublet into a cavity line $\omega _0+ \lambda^2/\Delta $ and 
an `atomic line' $\omega _0- \lambda^2/\Delta $.  
In atomic physics there are well established 
experiments~(Brune {\it et al.} 1990, Bernardot {\it et al.} 1992),
to detect such splittings.
In fact, detection of such lines is considered as 
a very efficient way of `quantum non-demolition' measurement~(Brune 
{\it et al.} 1991)
for small microwave photon numbers. Such atomic physics experiments, 
therefore,
may be used as a guide in 
performing the corresponding biological experiments 
involving MT (as cavities), as suggested in the present work.}.

Another aspect of our approach, if it turns out to be realized by brain 
cell MT,
concerns information processing and memory 
printing via {\it holographic} mechanisms. 
We have briefly discussed possible ways for the emergence of
holographic processing, which makes use of our quantum mechanical findings.
In particular, we have conjectured that quantum mechanical 
holography may be useful in transmitting information from the MT arrangements
to the synapsis of neuronic cells, and from there by {\it quantum 
tunnelling} to further-away regions. 
The hologram is argued to be produced in a situation analogous to the 
internal source X-ray holography in atomic physics. There, 
quantum electrodynamics is considered responsible for inducing the 
basic wave-interference equations that produce the hologram,
as a result of the interference of the quantum mechanical
paths between (virtual) photons inside the crystal. In the MT case, 
analogous interference phenomena occur as a result of the 
quantum superposition of the paths of the quantum coherent modes
in the ordered-water cavity interior, and/or in the dimer `lattice'.
The information was argued to be recorded in the presynaptic 
vesicular grid, from which, by quantal emission (`firing'), information 
is transmitted across the synapsis by quantum tunelling.

We believe that our analysis,
although speculative at this stage, however may lead to concrete
experimental 
set-ups involving MT, with the ability of testing 
possible quantum mechanical effects that may play 
a r\^ole 
on 
the processes of the transmission of electric signals (stimuli) 
by the neuronic systems, and even 
in the processes of memory, conscious perception, and decision making. 
Clearly much more work needs to be done before even tentative
conclusions are reached. However, we believe that the present work 
constitutes a useful addition to the programme of understanding 
the nature of the MT arrangements inside the cell, and the associated
processes of energy and information transfer across the cells. 
It goes without saying, of course, that one should be extremely cautious
when extrapolating results from {\it in vitro} (Laboratory) experiments,
involving Biological systems, to {\it in vivo} situations.

\section*{Acknowledgements}

This work is based on an invited talk by N.E.M. at the 
Workshop {\it Biophysics 
of the Cytoskeleton}, Banff Conference Center, 18-22 August 1997, 
Canada. We thank J. Tuszynski for his interest in our work. 
The work of N.E.M. is supported by P.P.A.R.C. (U.K.), and that 
of D.V.N. is supported
in part by D.O.E. Grant
DEFG05-91-GR-40633.

\section*{References}

\noindent  Agarwal, G.S., \underline{  Phys. Rev. Lett}, 
{\bf 53} (1984), 1732.

\noindent  Albrecht, A., \underline{  Phys. Rev.} {\bf D46} (1992), 5504.

\noindent  Bakas I.; Kiritsis, E., \underline{Int. J. Mod. Phys. A}
{\bf 7} (Suppl. {\bf 1A}) (1992), 55. 

\noindent Beck F; Eccles, J.C., \underline{Proc. Nat. Acad. Sci.}
USA {\bf 89} (1992), 11357. 

\noindent Bernardot F. {\it et al.}, \underline{  Europhysics Lett.} 
{\bf 17} (1992), 34.

\noindent Bjorken; J.D., Drell, S.D. \underline{Relativistic 
Quantum Mechanics} (Mc Graw Hill,
New York 1964). 

\noindent Brune, M. {\it et al.}, 
\underline{  Phys. Rev. Lett.} {\bf 65}(1990), 976.

\noindent Brune, M. {\it et al.}, \underline{  Phys. Rev.} A45 (1991), 5193.

\noindent Brune, M. {\it et al.}, \underline{  Phys. Rev. Lett.} 
{\bf 77} (1996), 4887.

\noindent Caldeira, A.O; Leggett, A.J., \underline{  Physica (Amsterdam)}
{\bf 121A} (1983), 587; \underline{  Ann. Phys.} {\bf 149}(1983), 374.

\noindent Del Giudice, E.; G. Preparata, G.; Vitiello, G.
\underline{  Phys. Rev. Lett.} {\bf 61} (1988), 1085.

\noindent Del Giudice, E.; Doglia, S; Milani, M.;
Vitiello, G., \underline{  Nucl. Phys.} {\bf B251 (FS 13)} (1985), 375;
{\it ibid} {\bf B275 (FS 17)} (1986), 185.

\noindent Dustin, P., \underline{MicroTubules}
(Springer, Berlin 1984);

\noindent Eccles, J.C., \underline{Proc. Roy. Soc. London} {\bf B227} (1986), 411.

\noindent Ellis, J.; Mavromatos, N.E.;
Nanopoulos, D.V., Phys. Lett. B293 (1992), 37;
\noindent {\it lectures presented at the
Erice Summer School, 31st Course: From Supersymmetry to the
Origin of Space-Time},
Ettore Majorana Centre, Erice, July 4-12
1993 ; Subnuclear Physics Series,  
{\bf Vol. 31} (1994), p.1 
(World Sci. );
\noindent For a pedagogical review of this approach see:
Nanopoulos, D.V., \underline{Riv. Nuov. Cim.} {\bf Vol. 17}, {\bf No. 10} 
(1994), 1.

\noindent Ellis,J; Hagelin, J.S; Nanopoulos, D.V.;
Srednicki, M., \underline{  Nucl. Phys.} {\bf B241} (1984), 381.

\noindent Ellis, J.; Mohanty, S.; Nanopoulos, D.V., 
\underline{  Phys. Lett.} {\bf B221} (1989), 113.

\noindent Ellis,J.; Mavromatos,N.E.;  
Nanopoulos, D.V., Proc. {\it 1st International Workshop 
on Phenomenology of Unification from Present to Future},
23-26 March 1994, Roma (eds. G. Diambrini-Palazzi {\it et al.},
World Sci., Singapore 1994), p.187.

\noindent Ellis,J.; Mavromatos, N.E.;  Nanopoulos, D.V., 
hep-th/9609238, \underline{  Int. J. Mod. Phys. } {\bf A}, in press; and
hep-th/9704169, \underline{  Mod. Phys. Lett.} {\bf A}, in press.

\noindent Engleborghs, Y., \underline{Nanobiology} {\bf 1} (1992), 97.

\noindent Fr\"ohlich, H., \underline{Bioelectrochemistry},
ed. by F. Guttman and H. Keyzer (Plenum, New York 1986).

\noindent Ph. Ghosez; X. Gonze; J-P. Michenaud, preprint {\bf mtrl-th/9601001} (1996), and references therein.

\noindent Gisin,N.; Percival,I., \underline{  J. Phys.} {\bf A26} (1993), 2233.

\noindent Gorini, V. {\it et al.},
\underline{  Rep. Math. Phys.} {\bf Vol. 13} (1978), 149.

\noindent Hameroff, S.R., \underline{  Am. J. Clin. Med.} {\bf 2} (1974), 163. 

\noindent Hameroff, S.R., \underline{Ultimate Computing}
(Elsevier North-Holland,
Amsterdam 1987);
\noindent  Hameroff, S.R.; Smith, S.A.;  Watt,R.C.,
\underline{Ann. N.Y. Acad. Sci.} {\bf 466} (1986), 949.

\noindent Hameroff S.;  Penrose, R.,
in \underline{ Towards a science of Consciousness}, The First Tucson
Discussions and Debates, eds. S. Hameroff {\it et al.} (MIT Press,
Cambridge MA 1996), p. 507-540.

\noindent Haroche S.; Raimond, J.M.,
\underline{Cavity Quantum Electrodynamics}, 
ed. P. Berman (Academic Press, New York 1994), p.123,
and references therein.

\noindent Jibu, M.; 
 Hagan, S.; Hameroff, S.R.; Pribram K.; 
Yasue, K.,  \underline{  Biosystems} {\bf 32} (1994), 195.

\noindent Lal, P., \underline{Physics  Letters} {\bf 111A} (1985), 389.

\noindent Lindblad, G., \underline{  Comm. Math. Phys.} {\bf 48} (1976), 119.

\noindent Mavromatos N.E.;  Nanopoulos, D.V.,  
\underline{  Int. J. Mod. Phys.} {\bf B11} (1997a), 851; 

\noindent Mavromatos N.E.;  Nanopoulos, D.V., quant-ph/9708003, 
\underline{  Int. J. Mod. Phys.} {\bf B12}, (1997b), 
in press. 

\noindent Mavromatos N.E.;  Nanopoulos, D.V.; Samaras, I.; Zioutas K.;
contribution to the Proc. 
of the Workshop \underline{Biophysics 
of the Cytoskeleton} (Banff, August 18-22 1997, Canada), to appear.

\noindent Miller.G.;  Sorensen,L.,  report \underline{DOE/ER/41014-4-N97},
cond-mat/9611155, and references therein. 

\noindent Nanopoulos, D.V.,  
hep-ph/9505374, 
Proc. \underline{XV Brazilian National Meeting} \\
\underline{on Particles and Fields}, eds. M.S. Alves {\it et al.}, 
Brasilean Phys. Society (1995), p. 28.

\noindent Nogales, E.; Wolf, S.;  Downing, K.,  
\underline{Nature}, {\bf 391 } (8 January 1998), 199.

\noindent Penrose, R., \underline{The Emperor's New Mind}
(Oxford Univ. Press 1989); \underline{Shadows of the Mind}
(Oxford Univ. Press 1994);

\noindent Pribram, K.H., 
\underline{Brain and Perception} (Lawrence and Erlbaum, New Jersey
1991).

\noindent Ricciardi L.M.; Umezawa, H., \underline{Kybernetik} {\bf 4} (1967), 44.

\noindent Sackett, D., presentation at the Workshop \underline{Biophysics 
of the Cytoskeleton}, Banff, August 18-22 1997, Canada.

\noindent Saito T.; Arimitsu, T., \underline{  Mod. Phys. Lett.} {\bf B7}
(1993), 1951.

\noindent Sanchez-Mondragon, J.J.; Narozhny, N.B.; 
Eberly, J.H., \underline{  Phys. Rev. Lett.} {\bf 51} (1983), 550.

\noindent  Satari\'c, M.V.; Tuszy\'nski, J.A.;
Zakula, R.B., \underline{  Phys. Rev.}  {\bf E48} (1993), 589;
\noindent this model of MT dynamics 
is based on the ferroelectric-ferrodistortive model 
of :  Collins, M.A.; Blumen, A.;
Currie,J.F.; Ross, J., \underline{  Phys. Rev.} {\bf B19} (1979), 3630;
{\it ibid.} 3645 (1979).

\noindent Y. Tsue; Y. Fujiwara, \underline{Progr. Theor.
Physics} {\bf 86} (1991), 443; {\it ibid} 469.

\noindent Tuszynski, J.A.; Brown, J.A., presentation 
at the \underline{Royal Society of London} \underline{Meeting} 
\underline{on Quantum Computation},
November 1997, 
London, U.K.

\noindent Tuszynski; J., Hameroff, S.; Satari\'c, M. V.; 
Trpisov\'a B.; Nip, M. L. A.;
\underline{J. Theor.  Biol.} (1995), {\bf 174}, 371.

\noindent Umezawa, H., 
\underline{Advanced Field Theory:} \underline{micro, 
macro and thermal concepts}
(American Inst. of Physics, N.Y. 1993).

\noindent Vitiello, G., \underline{Int. J. Mod. Phys. B} {\bf 9} (1995), 973. 

\noindent  Walls D.F.; Milburn, G.J., 
\underline{  Phys. Rev.} {\bf A31} (1985), 2403.

\noindent W. Zhong; D. Vanderbilt; K. Rabe, preprint 
{\bf mtrl-th/9502004} (1995), and references therein.

\noindent Zhu, Yifu, {\it et al.}, \underline{Phys. Rev. Lett.} {\bf 64} 
(1990), 2499.

\noindent Ziman, J.M., \underline{Principles of the Theory of Solids}
(Cambridge University Press, Cambridge, U.K.  1972)

\noindent Zioutas, K., presentation at the Workshop \underline{Biophysics 
of the Cytoskeleton}, Banff, August 18-22 1997, Canada. 

\noindent Zurek, W.H., 
\underline{  Phys. Today} {\bf 44}, {\bf No. 10} (1991), 36; 
\underline{  Phys. Rev.} {\bf D24} (1981), 1515.


\end{document}